\documentclass[12pt,preprint]{aastex}
\def\nad{\mbox {Na~I}}
\def\i{\mbox {\rm IRAS}}
\def\h1{\mbox {\rm HI}}
\def\x{\mbox {\rm X-ray~}}

\def\fig{{Figure}}

\def\deg{\mbox {$^{\circ}$}~}
\def\msun{\mbox {${\rm ~M_\odot}$}}

\def\lsun{\mbox {${~\rm L_\odot}$}}

\def\msunyr{\mbox {$~{\rm M_\odot}$~yr$^{-1}$}}
\def\msunyrk2{\mbox {$~{\rm M_\odot}$~yr$^{-1}$~kpc$^{-2}$}}
\def\msunpc2{\mbox {${\rm ~M_\odot ~pc}^{-2}$}}

\def\Ha{\mbox {H$\alpha$~}}
\def\Hb{\mbox {H$\beta$~}}
\def\Hg{\mbox {H$\gamma$~}}

\def\n{NGC~}
\def\asec{\ifmmode {'' }\else $''~$\fi}  
\def\amin{\ifmmode {' }\else $'~$\fi}    
\def\sles{\lower2pt\hbox{$\buildrel {\scriptstyle <}
   \over {\scriptstyle\sim}$}} 
\def\sgreat{\lower2pt\hbox{$\buildrel {\scriptstyle >}
    \over {\scriptstyle\sim}$}} 
\def\col{\mbox {\rm ~cm$^{-2}$} }
\def\kms{\mbox {~km~s$^{-1}$} }
\def\ergsec{~ergs~s$^{-1}$~}

\def\cm3{~cm$^{-3}$}
\def\cm5{~cm$^{-5}$}

\def\um{\mbox {$\mu${\rm m}} }

\def\et{{\rm et\thinspace al.}\ }   

\def\apj{ApJ}
\def\apjs{ApJS}
\def\pasp{PASP}
\def\aj{AJ}
\def\mn{MNRAS}

\def\aa{A\&A}

\begin{document}

\title{Mapping Large-Scale Gaseous Outflows in Ultraluminous Infrared
Galaxies with Keck~II ESI Spectra: Spatial Extent of the Outflow}

\author{Crystal L. Martin\altaffilmark{2,} \altaffilmark{3}}
\affil{University of California, Santa Barbara}
\affil{Department of Physics}
\affil{Santa Barbara, CA, 93106}
\email{cmartin@physics.ucsb.edu}


\altaffiltext{1}{Data presented herein were obtained at the W.M. Keck
Observatory, which is operated as a scientific partnership among the 
California Institute of Technology, the University of California and the
National Aeronautics and Space Administration.  The Observatory was made
possible by the generous financial support of the W.M. Keck Foundation.
}
\altaffiltext{2}{Packard Fellow}
\altaffiltext{3}{Alfred P. Sloan Research Fellow}

\author{Accepted to ApJ April 3, 2006}

\begin{abstract}

The kinematics of neutral gas and warm ionized gas have been mapped
in one-dimension across ultraluminous starburst galaxies using the \ion{Na}{1}
$\lambda \lambda 5890, 5896$  absorption-line  and \Ha emission-line profiles,
respectively, in Keck~II ESI spectra.  Blue-shifted, interstellar \nad\ absorption 
is found along more of the slit than anticipated, exceeding scales of 15~kpc in 
several systems.  If the outflow diverges from the nuclear starburst,  the absorbing 
gas must reach  similarly large heights along the sightline in order to cover 
so much of the galaxy in projection.  If the outflow is launched from a much larger 
region of the disk, however, then  the scale height of the absorbing material 
can be much lower. The large velocity gradient measured across a few outflows is, 
in fact, inconsistent with a flow diverging from the central starburst, as 
angular momentum conservation reduces the rotational velocity of an outflow as it 
expands.  In a few young mergers, where the galactic nuclei are still separated, 
the  spatial gradient of the outflow velocity is shown to be similar to that of 
the projected orbital motion across the system.  Widespread star formation 
triggered by the merger, or the galaxy-galaxy merger itself, may shock heat 
interstellar gas throughout the disk to generate a hot wind, in which  cool gas
presenting these properties is entrained.  The double nuclei systems present the highest outflow masses, mainly
due to the large area over which the cool gas can be detected in absorption against the
continuum.   In a typical ULIG, the mass carried by the cool phase of the outflow is at least 
$\sim 10^8$\msun,  which is a few percent or more of the total dynamical 
mass.  Assuming the starburst activity has proceeded at the observed rate for the past 
10~Myr, the kinetic energy of the cool outflows is a few percent of the supernova energy, 
consistent with starbursts powering the outflows.   The cool wind is expected to be 
accelerated by momentum deposition, possibly from radiation pressure as well as supernovae.
The turn-around radii of the cool outflows  are at least ~30 to 90~kpc, which presents a significant
\nad\ absorption cross section.  If most $L > 0.1L^*$ galaxies pass through a luminous starburst 
phase, then relics of cool outflows will create a significant redshift-path density.
Galaxy formation models should include this cool phase of the outflow in addition to
a hot wind in feedback models.
\end{abstract}


\section{Introduction}

Resonance absorption lines provide a more sensitive probe of diffuse 
gas in galactic halos than do emission lines.  The plethora of lines
in the rest-frame ultraviolet spectrum has made
it relatively straightforward to establish the ubiquity of outflows
in high redshift galaxies (Franx \et 1997; Pettini \et 2002; 
Shapley \et 2003).  Space-based observations show these lines are 
also blueshifted in the UV-brightest nearby galaxies (Heckman \& Leitherer 
1997, Kunth \et 1998, V\'{a}zquez \et 2004). Optical resonance lines 
also probe halo gas kinematics and allow measurements in more nearby galaxies;
but the paucity of lines limits these studies to low ionization states.
In particular,   interstellar
\nad\ $\lambda, \lambda 5890, 5896$ absorption is prevalent
along sightlines through the Milky Way halo and was recognized
in spectra of infrared luminous   galaxies (LIG's) 
(Armus, Heckman, \& Miley 1989). More recently, spectral observations
of this doublet have revealed cool outflows in many starbursts
(Veilleux \et 1995, Heckman \et 2000, Rupke \et 2002; Schwartz \& Martin 2004, 
Martin 2005; Rupke \et 2005b; Rupke \et 2005d).

One difficulty in the interpretation has been that the most often 
observed lines trace the kinematics of low-ionization species 
rather than the hot,  $ \sgreat 10^7$~K, gas that powers these 
multiphase outflows.  The abundant \nad\ lines probe regions where H 
is neutral -- i.e. relatively cool, dense, and dusty regions.
Hydrodynamic instabilities along the wind/disk interface, and at the
accelerating shell, generate fragments of interstellar gas that
are entrained in the hot wind.  The ram pressure of the hot wind 
probably accelerates these clouds (Heckman \et 2000, Fujita \et 2006),
although radiation pressure may also be important at ultraluminous
luminosities (Murray, Quartaert, \& Thompson 2004; Martin 2005;
Rupke \et 2005b; Rupke \et 2005d).  Martin (2005, hereafter Paper I) 
showed that some cool gas is accelerated to velocities $\sim 500$\kms\ 
in galaxies with $SFR > 10$\msunyr, but that the terminal velocities 
are lower in dwarf galaxies.  A simple dynamical model originally
suggested by Heckman \et (2000), and subsequently developed further
by Murray \et (2005),  was used in Paper~I to demonstrate that
the cool gas is generally accelerated to the terminal velocity of the 
hot wind by ram pressure; however, some dwarf starbursts do not supply 
enough momentum to bring the entrained mass up to the speed of the hot 
wind. Rupke \et (2005b) confirm this empirical result but suggest
the velocities saturate due to a lack of gas remaining in the path
of the outflow or a decrease in the efficiency of the thermalization
of supernova energy.


This paper presents longslit spectroscopy of the \nad\ absorption 
across ultraluminous infrared galaxies (ULIGs), defined as galaxies
with far-infrared luminosities similar to that of quasars --
$L_{Bol} \approx L_{IR} \ge\ 10^{12}$\lsun (Sanders \et 1988). ULIGs
may be the local analogs of the high-redshift, dusty starbursts that
contribute significantly to the cosmic star formation rate at $z \ge\ 1$
and dominate the far-IR/submillimeter background (Puget \et 1996).  The
presence of close nuclei and an overall disturbed morphology indicate
ULIGs are merging systems.  The collision leads to a central starburst
and the large-scale shocking of the gas disk (Jog \& Solomon 1992). The
relevance to this study is two-fold. First, the ambient medium that the 
starburst wind encounters is considerably more complicated than that 
encountered by the M82 or \n1569 winds because the tidal interaction 
propels streams of material outward and generates regions of inflow 
(Mihos 1999; Hibbard \et 2000; McDowell \et 2003; Colina \et 2004).
Second, the gas shock-heated by the merger may drive an outflow 
similar to that driven by supernova-heated gas in the starburst 
region (Cox \et 2004; Cox \et 2005).
  
Although spatially-extended, interstellar absorption has been detected 
in a few galaxies previously (Phillips \et 1993; Heckman \et 2000), 
the measurements described here provide one of the first systematic studies
of the angular extent and velocity gradients of the outflowing, absorbing
gas. This spatial information provides a  model-independent estimate of
the total mass in the cool outflow. Velocity gradients in the cool, absorbing gas are 
compared to the kinematics of the ionized gas  traced by \Ha 
emission, which helps  constrain the location of the absorbing gas along the sightline.  
Estimates of the mass-loss efficiency, momentum deposition, and
energy deposition are derived from these data. These results
provide insight into the appropriate dynamical model for the cool outflows
and empirically constrain feedback parameters in galaxy evolution models.
Due to the complicated, multi-phase nature of winds, a
dynamical model is required to understand the fate of the material in the outflow.

This paper is organized as follows. Section 2 describes the 
extraction and measurement of the longslit spectra. Section 3
presents the velocity gradients across the starbursts in
the \nad\ and \Ha lines.  The amount of material in the cool
outflow is discussed in Section 4. Implications for
the source of the outflow, the evolutionary products of ULIGs,
and the detectability of cool winds as intervening absorption-line 
systems are explored in Section 5. Section~6 summarizes the
conclusions.

\section{Observations}

Echellete spectra of 41 ultraluminous starbursts were obtained 2000 September 19-20 and 
2001 March 26-29 using ESI on Keck II (Sheinis \et 2002).   Galaxies were selected at 60 $\mu$m
from the IRAS  2 Jy sample (Strauss \et 1992). They have bolometric luminosities greater than  
$5 \times 10^{11}$\lsun, a 60\um\ bump  $F_{\nu}^2 (60\um) > F_{\nu} (12\um) 
\times F_{\nu}(25\um)$, and declination $\delta > -35\deg$.  The {\em CO-sample} contains 15 
ULIGs with CO velocities from Solomon et al. (1997) and 3 ULIGs with CO velocities from Dr. 
Aaron Evans (pvt. comm.).  The echellete has no automatic compensation for atmospheric dispersion 
and covers a broad bandpass ($\lambda 3900 - 1.1\um$ ), so the  position angle of 
the 20\asec\ ESI slit was generally chosen near the average parallactic angle. The exceptions 
were a few objects where structures in  R and K band images (Murphy \et 1996) warranted
specific position angles; and these systems were simply observed at a low airmass.
The spectral resolution, about 70\kms,  is not  high enough to resolve
individual clouds and shell fragments (Schwartz \& Martin 2004; Rupke \et 2005c;
Fujita \et 2006).

Paper I describes the extraction, reduction, and analysis of the one-dimensional nuclear spectra
in the CO subsample.  This paper focuses on the analysis of the two-dimensional longslit spectra. 
The reduction was identical to Paper I with the following additions. The longslit spectra were 
rectified independently for each order using arc lamp exposures. The sky was fitted along the 
spatial dimension and subtracted from the co-added data frames. The spectral images were
examined around \nad\ and \Ha.  A series of one-dimensional spectra were extracted to
enable measurements of the spatial gradients in the \nad\ line profile. Care was taken
to account for variations in plate scale among orders when extracting \Ha emission from the
lower order spectrum of the same physical region.  Each order was flux calibrated independently
using observations of spectrophotometric standard stars (Massey \et 1988).

\section{Kinematic Results}

The nuclear spectra presented for the CO-observed sample in Paper I reveal
cool gaseous outflows in 15 of 18 ULIGs. This incidence of winds, $\sim 80\%$ 
of ULIGs at $z \approx 0.1$, is higher than that in local infrared luminous
galaxies (Heckman \et 2000; Rupke \et 2005b). Since the systemic velocities are known
better for these ULIGs than others in the 2~Jy sample, this CO-sample is also used here 
to explore the gas kinematics across the merging galaxies. The cool component
of the outflow is traced in \ion{Na}{1} absorption, and the warm photoionized 
component is seen in \Ha emission. The absorption can occur at any location along
the sightline, but the peak of the \Ha emission profile will be dominated by denser 
gas, which likely resides in some form of a disk.  Random motions dominate
the stellar dynamics in ULIGs, but significant rotation is observed in the stellar
component; and the gas  presents even more rotation (Genzel \et 2001; Tacconi \et 2002).   
This decoupling of the gas and stellar dynamics occurs because of violent relaxation.
By the time the nuclei are 1~kpc apart, the stellar systems have nearly reached their 
equilibrium values of rotation and dispersion, but the accretion of high-angular momentum 
gas continues (Mihos 1999; Bendo \& Barnes 2000).

\subsection{Spatial Extent of Gaseous Outflows}

The \Ha\  emission (and more recently the \x emission) from galactic winds
have provided direct images of bipolar outflows from starburst galaxies
(Heckman, Armus, \& Miley 1990; Martin 1998; Veilleux, Cecil, \& Bland-Hawthorn 2005). 
In constrast to the absorption measurements, the faint halo emission is detected 
beyond the galaxy, usually along the minor axis. Indeed, the wind emission is hard
to see in projection against the galaxy because the disk is much brighter. The dominance
of the disk is illustrated by the 2D-\Ha spectra shown in the upper right panels
in Figure~\ref{fig:nad_spec}. 

The area covered by the ULIG outflows is explored in a systematic way here.
A previous (\nad) longslit measurement across the nearby luminous infrared galaxy, 
\n1808, Phillips \et (1993), detected \nad\ absorption on the near side of the 
nucleus and scattered emission off the back side. The slit spectra presented here
are ideal for examining outflow properties across the merging galaxies. The 
nuclei of the younger mergers yield independent high signal-to-noise sightlines, 
and the stellar continuum emission is sufficiently bright away from the nuclei
to measure gradients in the projected properties of any outflows.



\subsubsection{Extended \nad\ Absorption}

Figure~1 shows the \ion{Na}{1} spectra extracted at  (typically) five
positions, which are marked on  R band images kindly provided
by Lee Armus and Tom Murphy.  Two ULIGs with  strong nuclear \nad\ absorption, \i1056+24 and
\i1720-00, are shown in print; and the entire sample appears in the electronic edition. 
The width of the Doppler-shifted absorption component averages 
$<\Delta v_B> = 320\pm120$\kms FWHM toward the nucleus (Martin 2005), so 
the ESI data reveal a great deal of kinematic structure.  
Many of the off-nuclear sightlines  present complex line profiles, and the
line centers can be measured to $\sim 1 \kms$; the continuum S/N is adequate
 over about half of the 20\arcsec\ slit.
The absorption-line profiles for \i1056+24 are a remarkable example of
this kinematic structure. Toward the nucleus, Aperture~3 
in \fig~\ref{fig:nad_spec}, the absorption profile presents a 
high velocity tail. This component is also detected in Apertures 1 and 2
to the west. Surprisingly, the outflow extends beyond the main galaxy to the east.
It is detected against a companion galaxy in Aperture 5,  7.1 kpc (8\farcs4)
from the nucleus.  
Although the absorption equivalent width declines markedly over
this distance, the velocity of the line center is remarkably uniform.
If we picture the absorbing gas residing close to the disk (i.e. within a scale 
height), it is very difficult to understand why the projected bulk flow speed
would be this coherent over 11.4 kpc (13\farcs4).
The local star formation rate drops
significantly over this distance. Measured $U^{'}$ half-light radii for ULIGs are
typically around 5.3~kpc (Surace \et 2000).
Note also that this is the minimum angular extent of the absorbing gas; it cannot 
be detected at larger distances due to the faintness/absence of continuum emission.
In the context of an outflow diverging from the central starburst region,
the absorbing gas would have to be many kpc above the plane of the disk given 
the area covered by it. See \S3.2.3 for an alternative interpretation.

The size of the starburst region has been estimated from the infrared
luminosity.  The tight correlation of 100\um luminosity and CO luminosity
indicates the dust emission from ULIGs is optically thick at $\lambda < 
100$\um (Downes, Solomon, \& Radford 1993).  A minimum dust radius can 
therefore be estimated from a spherical blackbody at the measured dust
temperature (Solomon \et 1997); and the mean size of the starbursts
this paper is  $R \approx 200$~pc, reaching a maximum of 373~pc in \i1945+09.
The CO radii listed in Table~\ref{tab:ha}  are generally larger, presenting a mean of 320~pc and maximum
of 575~pc.  The starburst regions are clearly much
smaller than the region where outflowing gas is detected.


Table~\ref{tab:nad} summarizes the spatial extent of the absorption. For the
15 galaxies with cool outflows, the \nad\ doublet is detected over angles ranging from 
2\arcsec\ to 13\arcsec\ that subtend 4 to 18~kpc at the source. Mergers whose constituents 
are still visibly separated on the sky provide some of the widest sightline separations. Toward 
the object 7\farcs3 south of \i0018-08, the spectrum from Aperture 1 (in \fig~\ref{fig:nad_spec})
presents a marginal pair of \nad\ lines.
Unambiguous detections are readily apparent in the spectra of the close companions to
\i0015+54 (10.4 kpc), \i1150+13 (11.53 kpc), and \i1648+54 (5.41 kpc).
 The continuum S/N drops off before the absorption strength weakens in 
\i0803+52, \i1049+44,  \i1056+24,  \i1524+10, \i1720-00, and \i2008-03, so
the measured extent of the \ion{Na}{1} absorption is a lower limit for these galaxies.
 A few systems with extended continuum emission, from tidal features, do not present 
\nad\ absorption; these are \i1836+35 Aperture 1 (5\farcs7 S) and 5 (4\farcs6 N),
\i1929-04 Aperture 4 (8\farcs1 SE), and \i2336+36 Aperture 6 (6\farcs6 NW).


\subsubsection{Extended \Ha Emission}

The upper right panels in Figure~\ref{fig:nad_spec} show the 2D \Ha\ spectra along 
all the slits. The \i1056+24 spectrum illustrates some typical features of these spectra.
The \Ha emission comes predominately from one of the two merging galaxies; no 
\Ha emission is detected from the companion in Aperture 5. In contrast to the \nad\
absorption, the \Ha\ emission is confined to the center of the main galaxy.  The 
non-detection of extended \Ha\ emission from \i1056+24 in consistent with a face-on
outflow orientation, which is consistent with the morphology in HST NICMOS imagery 
(Scoville \et 2000). One expects near face-on orientations to favor absorption-line 
detections of winds, and that edge-on disks with winds would be the most likely to show 
extended \Ha and X-ray emission (Lehnert \& Heckman 1996; Strickland \et 2004a; Strickland 
\et 2004b). The gas velocity in the blue wing of the \Ha\ line profile is, however, often 
comparable to, and can exceed,  that of the most blue-shifted absorbing gas (Rupke \et 2005b).
Ionized gas is likely present in the bulk flows traced by \nad\ absorption, but the \nad\ and 
\Ha\ line profiles differ due to the quadratic dependence of the latter on the gas density
and the requirement that the former be in front of the continuum source.

The spectroscopic slit serendipitously intersected extended,
\Ha\ emission in 12 of the CO-selected systems. The spatial extent,
morphology, and velocities of the extended \Ha\ emission are summarized 
in Table~\ref{tab:ha}.  Of particular interest is a Doppler
ellipse discovered 8\farcs90 (21.2 kpc) north of \i0315+42, which
identifies an expanding shell of ionized gas. The blue- and red-shifted 
components of the line profile are separated along the northern half of the slit.
Using a shell radius of 5\farcs88 (14.0 kpc), which is the distance
the shell protrudes past the bright continuum of the disk, the dynamical
age and power requirements are large -- $\tau \sim 57$~Myr and
$L_w/n \sim 2.4 \times 10^{44}$\ergsec cm$^{3}$.\footnote{These values
  were estimated by assuming energy is continuously injected in a homogeneous 
  medium (Weaver \et 1977) .}
The power requirement is not uniquely determined because it depends on the 
density, $n$, of the surrounding material that is pushed out of the way.
A constant star formation rate of 1\msunyr\ and a Salpeter IMF from 1 to 
100\msun\ produces $L_{w} \approx 7.1 \times 10^{41}$\ergsec\ in mechanical
power from supernova (and stellar winds) (Leitherer \et 1999). Scaling to
\i0315+42, the maximum power is  $L_{w} \approx 2.1 \times 10^{44}$\ergsec.
Since the average density on these large spatial scales is likely $\sles\ 1$~cm$^{-3}$,
this energy injection rate appears sufficient to drive the expanding
shell.  No shell in visible in the continuum image. In the \Ha\ spectrum,
it is unclear whether the jet-like feature south of the nucleus is
the other lobe of a bipolar outflow because the line profile has a 
single peak.  These jet-like \Ha\ emission features are also seen
northeast of the higher velocity component in \i1049+44, east
of the higher-velocity component in \i1609-01, and toward the 
western side of the eastern component in \i1150+13.


The bipolar jets extending from \i2336+36 terminate at bright emission
blobs separated by 15.4~kpc in projection. Note that velocity shear
between the halo blobs is opposite to the direction of shear across 
the disks. The diffuse \Ha extending from each end of the galaxy
pair in \i0803+52 and \i1836+35 appears to be tidal in origin.
They are mirror images of each other but originate from respective
nuclei rather than a single starburst.  The smooth, diffuse halos
surrounding \i0018-08 and \i0026+42 could have a starburst-outflow
origin or a tidal origin.  The 2D spectral images dramatically show
that the extended emission does not share the disk kinematics in any 
of these cases. The excitation of this extended \Ha\ emission will
be discussed in a forthcoming paper.


\subsection{Velocity Gradients}

The \Ha emission is weighted toward the higher density gas near the disk 
because the emission measure scales as density squared -- $EM = \int n_e^2 ds$. 
The \Ha emission from most infrared luminous galaxies and a sizeable fraction
of ULIGs present rotation (Rupke \et 2005b).  Along the slit, the gradient
in the \Ha\ velocities at maximum intensity establishes a baseline for the
disk rotation to which the kinematics of the cool, neutral gas can be compared.   
For the CO-subsample,  \Ha spectra were extracted from the same physical apertures as the \nad\ 
spectra. The aperture locations are shown in \fig~\ref{fig:nad_spec}
on R band images (Murphy \et 1996)\footnote{Exact positions of
      each aperture can be determined from the slit position angle, see
      Table~\ref{tab:nad},  and the distance from the nucleus, shown
      in \fig~\ref{fig:pv}.}


\subsubsection{\Ha\ Kinematics}

The \Ha\ spectra show that the relative velocity of the merging 
galaxies helps distinguish the two nuclei better than imaging alone. 
Murphy classified 7 
of the 18 galaxies in the CO-sample as double nuclei based on
imaging with resolution $\sim 0\farcs8$ (or 1.48~kpc at z=0.1). The
spatial resolution of the Keck ESI data are similar, but 
the rotation of the individual disks make it possible to separate them
in two additional cases. The estimated separations are 
\i1150+13 (5\farcs0, 11.4 kpc) and \i1720-00 (3\farcs0, 2.5 kpc).
Genzel \et (2001) also classified \i1720-00 as a single nuclei system.
Their measurement of the stellar kinematics, $v_{rot}/\sigma = 0.48$,
is consistent with the mean of their double nuclei control sample,
however; so the reclassification suggested here is dynamically possible.
In general, the nuclear separation is a useful indicator of the dynamical 
age of a merger (Murphy \et 2001). That nine of the 18 ULIGs discussed here 
present double nuclei indicates half the sample have not advanced to 
the final stage of the merger yet.


Voigt profiles were fitted to the \Ha\ lines in \fig~\ref{fig:nad_spec},
and Doppler shifts were derived from the central wavelength. The \Ha
position-velocity (P-V) diagrams are shown (by solid circles) in 
Figure~\ref{fig:pv}.  The spatial origin is defined as the nucleus 
of the ultraluminous galaxy.  Tilted lines reveal projected rotation. 
Again using \i1056+24 for illustration, the \Ha\ P-V diagram reveals
the ionized gas in the galaxy is receding (approaching) on the 
western (eastern) side of the galaxy.  The velocity difference is
120\kms.  The low projected rotation speed of 60\kms\ is consistent
with a disk seen nearly face-on.\footnote{For example, a galaxy with circular 
        speed of  250\kms\ inclined 14\deg\ will produce the observed 
	 gradient.}
The nucleus lies close to the
region where the P-V curve crosses zero velocity. Indeed, the
rotation curves provide another check on the assigned systemic 
velocities.  The uncertainty in measuring the strong \Ha\ line 
profiles is neglible, but the positions in \fig~\ref{fig:nad_spec}
are a bit {\it fuzzy} for the following reasons. First, within each 
aperture, the position of the brightest region is not known. A 
shift perpendicular to the slit could shift the velocity scale by 
as much as 38\kms.  More importantly, the brightest 
emission knot could be anywhere in the aperture along the slit, so
the spatial error bars are half the separation between points.
And, finally, the heavy extinction toward these nuclei means that
the {\it nucleus} identified in the R band images 
may be slightly offset from the true nucleus at longer wavelengths.
Hence, as long as the velocity of the R-band nucleus is close to 
zero, there is no reason to doubt the systemic velocity.


In the position-velocity diagrams or two-dimensional \Ha\ images, the slope of the emission
across a single galaxy shows the projected direction of the 
rotational spin vector. Similarly, the tilt of an imaginary line 
connecting the centroids of the \Ha\ emission of the two galaxies
indicates the projected orbital rotation. For example, in Apertures 1, 
2, and 3 for the \i1720-00 pair, the \Ha\ velocity shear traces 
the rotation curve of the northern galaxy. 
Apertures 4 and 5 show the projected 
rotation of the southern galaxy.  Notice that the velocity shear 
between the two galaxies in \i1720-00 is of the same order as 
that within the northern galaxy itself. Hence, the \Ha\ PV-diagrams 
reflect orbital kinematics as well as rotational motion.

\subsubsection{\nad\ Kinematics}

Spectra around the \nad\ doublet were extracted from the same locations as 
the \Ha\ spectra.  These apertures, which are marked in Figure~1,  were chosen 
to optimize sensitivity to interstellar absorption.  As in Paper~I (Martin 2005), 
the simplest fitting procedure holds one pair of lines at the systemic velocity and finds 
the best-fit velocity for a second pair of lines. The absorption troughs in ULIG
spectra are more complex than this simple description, but this approach has the
advantage of characterizing the bulk velocity 
of the dynamic component without introducing a model-dependent components. The fitted
velocity depends on the doublet ratio, which was initially fitted. As described 
in \S 4.0, constraining the column densities required a more sophisticated approach;
and these  spectral fits are shown in \fig~\ref{fig:nad_spec}. The velocities measured 
with both methods are plotted in Figure~2 and made available in Table~3.
These approaches generally yield consistent velocities, particularly when the absorption 
troughs present significant velocity structure.  It was important, however, to account
fot the systemic absorption component, which was sometimes much stronger than that from stellar
absorption  alone (see Paper I). The \nad\ fitting was carried out with the IRAF
distribution of specfit (Kriss \et 1994). Limitations of the specfit parameter
range are discussed by Rupke \et (2005a). In the small number of cases where difficulties
were encountered in practice, fitting parameters where frozen at a plausible value and
flagged. The spatial resolution of the data presented here, typically $\sim 1 - 3 ~{\rm ~kpc}$, 
provides several velocity  measurements across each ULIG, which are used here to 
explore velocity gradients across the outflowing, absorbing gas.

The \nad\ absorbing gas is not found in simple rotation like much of the emission-line
gas. Across \i1056+24, all the \nad\ velocities are blue-shifted relative to the systemic 
velocity. The bulk of the absorption trough is rarely redshifted.  One unique exception 
is \i0803+52, where all interstellar absorption is redshifted; and the cool gas is either 
infalling or ejected from a foreground component of the system. Section~3.2.3 and 
Appendix~A explain how outflows could produce redshifts at a few positions.  In systems 
like \i1150+13,  redshifts can also be produced if an extended rotating disk
is seen in projection against a galaxy behind it. The position -- velocity plots 
in \fig~\ref{fig:pv} clearly  demonstrate, however, that the bulk of the absorption trough 
is generally blueshifted relative to the CO (and \Ha) velocities.

The left panel of Figure~\ref{fig:velgrad} compares the velocity gradients measured across 
all the systems in \Ha\ and \nad; these measurements are summarized in Table~\ref{tab:ha}.
Many of the galaxies, including \i1056+24, present relatively 
small \nad\ velocity gradients, falling below the diagonal line in \fig~\ref{fig:velgrad}.  
Galactic winds are not expected to exhibit much rotation.  In most models,  outflows diverge 
from the central region of a galaxy, so angular momentum conservation slows down any rotational 
component present at launch.  Appendix~A describes a simple geometrical model of such
an outflow and demonstrates that the projected velocities should be nearly constant along 
a slit aligned with the major axis of the galaxy. Panel {\it a} of \fig~\ref{fig:vgrad_na} shows 
several systems with little velocity gradient -- \i1524+10, \i1929-04, and \i2336+36.
Higher resolution mapping of the gas distribution is required to determine whether the slit 
position angle follows its major axis.

%

A significant velocity gradient was discovered across \i1720-00. In the \nad\ longslit 
spectrum (shown in \fig~\ref{fig:nad_spec}), the gradient is visible in the absorption lines 
between apertures 1 and 5. The position-velocity diagram shows that
the \nad\ lines shift by 42\kms~kpc$^{-1}$ across the system, which is just as large
as the \Ha velocity gradient of 47\kms kpc$^{-1}$.  Considering the large 
offset, about 400\kms, 
between the \nad\ absorption and the \Ha emission, the \nad\ and \Ha\ velocity gradients 
are remarkably similar.
In both \nad\ absorption and \Ha\ emission, the projected orbital motion 
of the merging galaxies produces much of this shear. Tic marks in the position-velocity 
diagram mark the spatial extent of each galaxy along the slit. The velocity
gradient between the nuclei of \i1720-00 is  38\kms~kpc$^{-1}$. For all the double nuclei
systems that had line detections in both galaxies, this orbital shear was measured
using the \Ha\ emission line. The right panel of \fig~\ref{fig:velgrad} shows 
that the total \nad\ velocity gradient across these is similar to the orbital
shear. The cool outflow in \i0015+54 shows similar kinematics;  i.e. the
\nad\ absorption has a large blueshift relative to the \Ha\ emission, but
both have large velocity gradients that are partly due to orbital motion.
The absorbing gas cannot be part of the disk in these two systems. A rotating
gas disk would produce redshifted absorption on one side of the center of rotation.  
The apparent rotation could be explained by invoking separate outflows from each nucleus. 
However, the starburst activity is often much stronger in one nucleus, so it is
not obvious that two outflows are produced.


Some cool outflows present more rotation  than expected.
Across \i1720-00,  apertures 1 and 2 cover the ULIG, and apertures 3 and 4 
cross the companion. The P-V diagram shows the rotation curve of the ULIG
is 65\kms~kpc$^{-1}$, while the companion appears to rotate in the 
opposite direction at 20\kms~kpc$^{-1}$.   A similar paradox has been 
described for halo gas in the spiral galaxies \n891 and \n5775. The gas 
velocity falls off more slowly with scale height in \n891  and 
(to a lesser extent) in \n5775  than ballistic models predict (Rand 2000, 
Tullmann \et 2000). An outward pressure gradient, or drag between the 
disk and halo, seems to be required (Collins \et 2002). The situation is 
not completely analogous across \i1720-00 and  \i0015+54 because much of 
the velocity gradient arises from orbital motion.  Rotating halo gas is also a 
common property of intermediate redshift ($z \sim 0.5$) spiral galaxies (Steidel \et 2002). 

Six systems in the left panel of  Figure~\ref{fig:velgrad} present a
\nad\ velocity gradient that is larger than the \Ha\ gradient.  In only two
systems, \i0018-08   and \i1049+44, are the velocities not constrained well 
enough in  for the large \nad\ gradient to be significant.
For pre-mergers, the velocity gradient due to the projected rotation 
of the galactic gas was measured from \Ha velocity gradient across the ULIG member 
of the galaxy pair.  The middle panel of Figure~\ref{fig:velgrad} compares
the total \nad\ velocity gradient to this rotational gradient.  The similar
loci of the data in the middle and left panels indicates that much of the \nad\ velocity 
gradient reflects rotation of the galactic gas. When the slit is aligned along 
the minor axis of the outflow model from Appendix~A, high inclination
observations produce a velocity gradient across the center of the galaxy. 
Four systems are shown in panel {\it b} that could plausibly reflect this 
type of break in their P-V diagrams. The absorption velocities become 
redshifted when the outflow cone is pointed away from the observer
on the far side of the minor axis.  For the case of \i1150+13, an alternative
explanation is that the receeding western side of the gas disk associated with 
the eastern galaxy is being detected in absorption against the western galaxy. This geometry
naturally explains the redshift, but  the \nad\ column densities do not
decrease as rapidly as might be expected for a disk seen in projection.
Another origin is needed for the rotation 
in \i0315+42 and \i1836+35 because the \nad\ velocity
gradient is in the opposite sense of that of the star-forming disk.\footnote{
         The gradient is unlikely minor axis shear. The \i0315+42 system
	 is thought to be viewed edge-on in light of the detection of the 
	 \Ha shell, but it would need to be viewed at low, 
	 $ i < 90\deg - \theta_w$, inclination to keep the 
	 velocities on both sides of the nucleus blueshifted as observed.}

Spatially-extended starburst activity throughout both gas disks would explain the oberved
rotation of the outflow. If the the wind is generated over a large fraction of the disk,
then the absorbing gas need not be at extremely large heights to completely cover
the continuum source. Higher resolution studies of the star formation activity are
needed to confirm the viability of this explanation, however. The size of the
molecular gas disk and the far-infrared emission are much smaller than the
area covered by the outflows described in this paper.


%


\section{Mass of Cool Gas in Outflow}


Measurements of mass outflow rates are important for understanding
the influence of winds on galactic evolution and the intergalactic
medium.  The mass of hot and warm gas in winds has been shown to increase 
with the SFR (Martin 1999).  The total mass in the cool component is
difficult to estimate.  The strength of the \nad\ absorption along the slit,
and its relation to the amount of cool gas along the sightline, are 
discussed here.


\subsection{\nad\ Column Densities}

Extraction of the \nad\ column density is complicated by the 
blending of the $\lambda \lambda 5890, 5896$ lines and their
saturation.  The inferred columns grow non-linearly with the
line strength in this regime, so seemingly minor degeneracies
in profile fitting can lead to enormous allowed ranges for 
the estimated columns. A comparison of standard curve of growth 
techniques (Spitzer 1968) and more specialized methods was 
therefore required.  For a related but distinct approach to fitting
absorption troughs in ULIG spectra, see Rupke \et 2005a.

The large width of the \nad\ absorption trough in ULIGs indicates
the presence of multiple absorbing regions along the sightline; 
see, for example, the simulations of Fujita \et (2006). The 
underlying velocity distribution is likely more complex than 
the Gauassian profile assumed in the standard curve-of-growth method. 
Simple deconvolution techniques were therefore intially applied to
the line profiles using a kernel composed of two narrow lines of
fixed intensity ratio, which are separated by 300\kms. The resulting 
templates had Gaussian-like features including, for
example:   (1) A single Gaussian centered at the
systemic velocity, (2) A smooth distribution with a maximum at the
systemic velocity but a strong skew toward blueshifted velocities,
and (3) two Gaussian profiles - one at the systemic velocity and
one Doppler shifted. The ambiguity in the kernel can, in
principle,  be eliminated by working on the apparent optical depth 
profile instead of the intensity profile, as $\tau_{5890}(\lambda)  = 
\tau_{5896}(\lambda + s) (f \lambda)_{5890} / ( f \lambda)_{5896} $.
Substituting for the separation of the doublet $s$ and the oscillator
strengths,  this expression becomes
$\tau_{5890}(\lambda)  = 1.98 \tau_{5896}(\lambda (\AA)  + 6.97 (1  + z))$. 
In practice, the optical depth is only
a simple function of the intensity profile when overlapping 
absorbers completely cover the continuum source.  
The constraint imposed by the doublet can also introduce oscillations 
in the solution (Arav \et 1999a), and regularization methods have not been 
developed for the case of partial covering.  


Since the templates suggested by this exercise are well-behaved,
pairs of Gaussian profiles were fit to the spectra following  Heckman 
\et (2000), Rupke \et (2002), and Martin (2005). When the underlying template 
can be described by many Gaussian components, and the invidual components 
are well-behaved -- i.e. $\tau_0 \sles\ 5-10$ and the Doppler parameters 
are not varying wildly, multiple components can be analyzed collectively 
using standard, single-component curve-of-growth methods (Jenkins 1986).
In Paper~I, one component was required to be at the systemic velocity;
and outflow/inflow was associated with one (or more) additional components
as required to described the overall profile shape.  This approach
constrained the doublet ratio well provided the line profiles had multiple 
absorption troughs; but the doublet ratio had to be guessed for systems
presenting smooth absorption troughs. The doublet ratios listed in
Table~3 of Paper~I imply very high optical depths, $\tau_0 > 10$,
for 6 systems  -- \i0015+54, \i0018-08, \i1049+44, \i1056+24, \i1720-00, 
and \i1836+35.
(Here, $\tau_0$ is the estimated optical depth at the center of the 
$\lambda 5896$ line.)  

%

The curve-of-growth method clearly breaks down when the lines are strongly 
saturated.  The accuracy of alternative methods, like integrating the 
apparent optical depth, are also limited by the degree of saturation.  
The spectral resolution of 70\kms may not resolve individual components,
and the only way to ensure accurate measurements is to obtain much higher 
resolution spectra in the hope of resolving out the saturated regions. 
For now, one self-consistent approach is to measure a lower limit for
the optical depth and then fix the doublet ratio to give this same optical 
depth. 

In practice, extraction of the apparent optical depth is complicated
by the non-unity covering fraction of the absorbing gas.
High-resolution, ultraviolet spectra of both high-redshift and 
nearby starbursts often present completely black, low-ionization 
lines (Pettini \et 2002; V\'{a}zquez \et 2004).
Because the doublet ratio indicated the \nad\ 
lines are optically thick, the non-zero residual intensity can be
attributed to partial covering of the continuum source.
A single aperture averages sightlines over areas
of a square kiloparsec, which are likely much larger than the dense regions 
where \ion{Na}{1} is abundant. Along sightlines through our Galaxy,
the \nad\ doublet reveals  small scale 
 structure, around 3200~AU (e.g. Andrews, Meyer, \& Lauroesch 2001). 
The difference in covering
factor between the optical and UV lines may simply reflect the patchiness
of the UV light and its association with outflow regions.  When the
doubet lines are not blended, the covering factor can be derived directly
from their residual intensities (Barlow \& Sargent 1995; Hamann \et 1997).
To extract the optical depth from the intensity profile, a velocity-independent
covering factor was simply estimated from the residual line intensity as
$C_f \approx 1 - min(I(\lambda))$. 
The average value for the nuclear apertures  are $C_f \sim 30\%$ 
(Martin 2005). If the overlapping line components are ignored, and
the $I_{5890}$ measured from the bottom of the observed profile,
the mean increases to $C_f \sim 45\%$;  these values are used here
for comparison to Heckman \et (2000).  Galaxies with larger
covering factors nearly always show stronger \nad\ lines (Heckman \et 2000;
Martin 2005). Inspection of \fig~\ref{fig:nad_spec} shows that the nuclear 
covering factor varies among galaxies, and that the covering 
factor typically decreases with distance from the nucleus along any 
particular slit.

An apparent optical depth
profile,  $\tau(\lambda)$, was extracted from each spectrum.  When
multiple, overlapping absorbers are present, the intensity of the
absorption trough, $I(\lambda)/I_0 = C_f e^{-\tau} + 1  - C_f$, is 
determined by the amount of light not covered by absorbers, $ (1 - C_f)I_0$, 
and the optical depth in each transition at any wavelength, 
$\tau(\lambda) = \tau_{5890}(\lambda) + \tau_{5896}(\lambda)$.  
The integrated optical depths 
obtained with this procedure lie in a fairly narrow range,
$2 < \tau_{5895} < 10$.
Because upper limits were adopted for the absorber covering fraction, 
the estimated columns may be systematically low. 
This decomposition is the simplest one consistent with the data.
See Rupke \et (2005a) for a description of other possible absorber
geometries.




The optical depth of the \nad\ $\lambda 5896$ transition was used
to estimate the column density of the ion. The {\it apparent}
column density,
\begin{eqnarray}
  N_{ap} = \frac{m_e c^2}{\pi e^2 f \lambda_{0}^2} 
  \int \tau_{\lambda} d\lambda,
\end{eqnarray}
or
\begin{eqnarray}
  N_{ap}(NaI) = 1.02 \times 10^{13} {\rm ~cm}^{-2} 
\int \tau_{5896}(\lambda) d\lambda
\end{eqnarray}
for the $\lambda 5896$ transition, is a lower limit on the true column.
The results are given in column 6 of Table~3.  They generally
agree, to within a factor of a few,  with  the column densities obtained
by fitting the intensity profile, as shown in column 8 of Table~3 .
Four of the ULIGs are included in the low-redshift 
Rupke \et (2005a) sample. The estimated columns for the nuclear aperture 
are within a factor of two of their values, which is very good 
agreement considering the different fitting techniques, possible aperture 
differences, and nonlinear response of the columns.  As a cautionary note,
columns 11 and 12 show the minimum and maximum column densities consistent
with the intensity profiles. These extrema were obtained by fitting the 
intensity profile and allowing the doublet ratio to vary, and they
illustrate the potentially large uncertainties that arise from the
non-linear relationship between the line profile and the optical depth.
In addition, it should be recognized that the absorption troughs represent 
averages across the continuum source. Since the non-linear weighting strongly 
favors low optical depth paths, even techniques that fit the optical depth will 
tend to underestimate the column density.




\subsection{Estimated Mass in the Cool Wind} \label{sec:mass}

Estimating the H gas column $N(H)$ from the \ion{Na}{1} column
requires corrections for the abundance of Na relative to H,
the depletion of Na onto dust grains, and the amount of ionized Na.
Sub-solar abundances measured along sightlines through the disk and halo 
of the Milky Way have been interpreted as depletion onto dust (Wakker \& 
Mathis 2000; Savage \& Sembach 1996); and a solar Na/H abundance ratio, 
$2.04 \times 10^{-6}$ (Morton 1991 and references therein), is adopted 
for these massive  ULIGs. A conservative conversion to total HI column,
\begin{eqnarray}
 N(HI) = 4.90 \times 10^{20} {\rm ~cm}^{-2} 
\left( \frac{N(NaI)}{10^{14} {\rm ~cm}^{-2}} \right) \times \nonumber \\
\left( \frac{ f_d  (N(Na)/ N(NaI)) }{ 10} \right),
\label{eqn:nh}  \end{eqnarray}
is applied here, where $f_d$ is the depletion factor $d_{Na} = 
d_{X} \equiv (X/H)_{\odot} / (X/H)$.  
From the range of measured N(\ion{Na}{1}) in Table~3, the
corresponding H columns are estimated to range from  $N(H) \approx
8.8 \times 10^{19}$\col\ in \i1056+24 (aperture 4 component 1) to as much as 
$N(H) \approx 1.2 \times 10^{21}$\col\ in \i0026+42 (aperture 1).
In many regions where H is neutral, 
NaI is probably not the dominant ionization stage. (The  ionization 
potential of sodium, 5.1~eV,  is much less than that of HI.) In the Milky Way, 
the product of the Na depletion onto dust grains and the \ion{Na}{1} 
ionization fraction presents little variation with N(H). Provided the 
H column is less than $10^{21}$\col, their product  is $\sim 300$,
within a factor of a few (Wakker \& Mathis 2000). Since the correction may 
decrease by an order of magnitude at higher columns due to self-shielding of Na from 
ionizing  radiation (Ferlet \et 1985), a correction factor $d_{N~I} d_{ion} = 10$
is adopted here for ULIGs to avoid overestimating the mass column. 
Using the mean values for intermediate velocity clouds in the Milky Way 
would raise all masses quoted here by a factor of 30. Note that
Schwartz \& Martin adopted $d_{NaI} d_{ion} = 100$ in their study of 
dwarf starburst galaxies because they have significantly lower \nad\ columns.

The run of column density along each slit constrains the total mass of 
the cool outflow better than the nuclear sightline alone.  To estimate the 
outflow mass, the column density at a particular position was multiplied by 
the area of an annulus, centered on the nucleus, and the estimated gas
covering fraction. The radius was set  equal to the aperture's distance from 
the nucleus, and the thickness of the annulus determined by the aperture 
spacing. The masses from concentric annuli were added to arrive at the total mass
of cool gas in the outflow.   A significant fraction of the total mass is detected 
in the off nuclear sightlines.
In several galaxies, the absorption
is strong out to the limit of the continuum detection, so the areas
are lower limits; and masses are given as lower limits in Table~\ref{tab:nad}.

Figure~\ref{fig:m} shows the mass of the outflow derived this way.
Nearly all the error ranges are consistent with $M_c \sim 6 \times 10^7$\msun\ 
in the outflow.  Although no simple relation with the star formation rate
is observed, the baseline in SFR is not large; and the lack of correlation
is not surprising.   Similar conclusions about the cool wind mass 
in ULIGs have been reached by fitting free-streaming outflow models (constant
velocity and mass flux) to nuclear spectra (Rupke \et 2005b).
The mass estimates made here contain conservative assumptions about line 
saturation and the Na ionization/depletion correction. It is therefore
plausible that the masses are actually systematically higher than shown.
In addition, if the mass is calculated using the the extreme limits on the 
column density, or $f_df_{ion} \sim 100$, then 
 values of $\sim 10^9$\msun\ are allowed in many ULIGs. 

Figure~\ref{fig:mdyn} shows that the cool outflows carry a significant amount of 
mass. The molecular gas mass ranges from $4 \times 10^9$\msun\ to $15 \times 10^9$\msun\ for 
these galaxies (Solomon \et 1997), and the interstellar gas is predominantly molecular 
in the ULIG phase.  The mass inferred for the cool outflows are typically a few percent of the 
dynamical mass in the starburst region, where $R_{CO} \approx few \times 100$~pc.
The \nad\ column densities measured toward dwarf starburst galaxies are lower than those
in ULIGs, and the outflow speeds are also lower (Schwartz \& Martin 2004). Using
the atomic gas mass referenced by Schwartz \& Martin, and adding molecular
gas masses in the central regions (Hunter \et 1996, Scoville \et 
1991, and Wei\ss\ \et 2001) yields the points for dwarf starbursts
shown by the open symbols in Figure~\ref{fig:mdyn}.  The dwarfs 
present lower columns of blueshifted \nad.  Even when larger ionization
corrections are assumed, e.g. $f_d f_{ion} \approx 100$ in Schwartz \&
Martin (2004), the mass of outflowing gas is considerably smaller in
dwarf starbursts than it is in ULIGs. The
fraction of the interstellar gas mass in the cool wind is similar
to that measured in the ULIGs. The interstellar gas masses for the 
dwarfs may be overestimated in that some of the  HI gas is not
associated with the starburst region, and this correction would
shift the outflows in dwarf starbursts to higher gas fraction. The  galaxy that appears to lie 
at 100\% efficiency is \n4449; and the wind mass is likely an overestimate
as this component was the most saturated in the Schwartz \& Martin (2004)
sample.



\subsubsection{Dust Associated with the Cool Gas}

When optically thick gas only partially covers the continuum source, the \nad\ 
line strength is sensitive to the velocity spread of the absorbers and their 
covering factor (in addition to the optical depth).  Previous work has shown
that the interstellar \nad\ strength in starbursts is more correlated with 
covering factor than with line width (Heckman \et 2000; Martin 2005). Since
the reddening also tends to be larger toward luminous infrared galaxies with strong 
\nad\ absorption lines (Armus, Heckman, and Miley 1989,  Veilleux  \et 1995), 
Heckman \et (2000) suggested that more reddened sightlines have larger covering 
factors of cool gas and dust. For the ULIGs discussed in this paper, the reddening
was estimated using the Balmer decrement; and Table~\ref{tab:ha} lists the results.
The measured \Hb/\Hg flux ratio consistently indicated higher reddening than did 
the \Ha/\Hb ratio; but the discrepancy is attriubted to underestimating the strength
of the broad, stellar absorption at \Hg, which was fit simulataneously with the 
emission component.  Figure~\ref{fig:bv} confirms that the reddening tends to
be larger in ULIGs with strong \nad\ absorption. Of the 5 highest equivalent width 
ULIGs, the covering factor is high in four systems -- \i1056+24, \i1524+10, 
i1720-00, and \i1836+35. The absorption in the other galaxy, \i2008-03, presents an
average covering factor but the largest velocity width in the entire sample.
\fig~\ref{fig:bvcf} shows that all the galaxies with  high covering factor are  
indeed highly reddened. However, some highly reddened starbursts have average 
\nad\ covering factors. The ULIGs and LIGs present the same range of covering 
factors, but the ULIGs have higher average reddening.

%



It is interesting to examine how much of the reddening might be associated 
with dust in the wind.  Along sightlines through the solar neighborhood, the 
amount of reddening is correlated with the gas column (Spitzer, \fig 1.1; Savage \et 1977). 
Extrapolating this Galactic gas-to-dust ratio to the outflowing gas columns 
in ULIGs yields considerably less reddening than is measured.   Even for the 
sightlines with the largest columns, $N(NaI) \approx 10^{14}$\col, the
correction factor $f_d f_{ion}$ would have to be much larger than 10
for all the dust to be associated with the the outflowing gas column.
It is not surprising that the reddening is larger than that associated 
with the outflow because there is a signifiant gas column at the systemic 
velocity; and this is consistent with the variation in \nad\ covering
factors among highly reddened ULIGs and perhaps the offset between LIRGs
and ULIGs.

%






\subsubsection{Relation of Wind to Merger Process}


The estimated outflow masses are more uncertain than the column densities
because the area covered by the absorbing gas is not fully constrained.
The \nad\ doublet can be detected at larger radii in dynamically younger
mergers because these mergers have wider separations (larger continuum baseline).
Four of the five galaxies with mass estimates above $10^8$\msun\ are
double nuclei systems. Six of the eight double nuclei systems, \i0015+54, \i0018-08, 
\i1150+13,  \i1056+24, \i1720-00, and \i2008-03, have large estimates for the outflow
mass, upper limits of $\sim 10^9$\msun.  (For the other two double nuclei systems,
\i0352+00 and \i1648+54, no \nad\ absorption was measured toward the nucleus.)  
The extent of the absorbing gas is at least 5  to 18~kpc in the double nuclei 
systems, whereas absorption is detected in the later stage mergers to a radius of 
only 4 to  7~kpc.  The single nuclei systems, particularly those  without obvious tidal features --
\i0315+42, \i0803+52, and \i1929-04 in particular, do appear to have lower mass outflows.
However, the difference in estimated mass arises entirely from the apparently smaller
area covered by the outflow, as the column densities toward the nuclei do not differ
appreciably (in moderate spectral resolution at least). Rupke \et (2005b) also find
no statistically difference among outflows in pre-mergers, mergers, and old mergers.

The separation of galactic nuclei and tidal features are good indicators of the relative 
dynamical ages of ULIGs (Murphy \et 2001). Outflows are expected to evolve with the progression 
of the merger. Galaxy-galaxy mergers likely trigger multiple phases of star formation, starting 
with a nuclear starburst but later progressing across the disk (J. Barnes, pvt comm). 
Nuclear activity is also expected to be exposed as these mergers advance  (Hopkins \et 2005).
Measuring the evolution in the outflow column density is complicated, however, by the
sensitivity of the sodium ionization corrections to the hardness of the radiation field
(Murray \et 2006).  Although the only classified Seyfert nucleus in the CO-subsample is 
\i1150+13, several of these ULIGs have LINER-like optical line ratios.  Overall, the column
density measurements are not precise enough to explore such  variation.  However, 
ULIGs with clear signs of nuclear activity do show systematically higher outflow velocities
(Martin 2005; Rupke 2005d). Also, ULIGs with starburst/AGN-composite spectra show
some evidence for a contribution to the momentum from radiation pressure  (Rupke \et 2005d,  
Fig. 12b discussion).


\section{Discussion}

Galactic winds are ubiquitous in starburst galaxies, and the
mechanical power supplied by temporally (and spatially) correlated
supernovae is sufficient to power them (Heckman \et 1990).   The 
empirical correlation between the mass loss rate of {\it warm}/{\it  hot} 
gas and the star formation rate supports this scenario (Martin 1999). 
The detection of neutral gas in starburst outflows came as a surprise
(Heckman \et 2000; Rupke \et 2002; Schwartz \& Martin 2004; 
Martin 2005; Rupke \et 2005b; Rupke \et 2005d), because it is not 
obvious that such dense clouds survive in a wind. This paper takes
a closer look at the cool component of outflows in ULIGs.  In \S~\ref{sec:dynamical},
it is argued that the mass loss rates are significant but may not be as large a 
fraction of the SFR as was estimated for the warm and hot phases (Martin 1999), a point 
made previously by Rupke \et (2002, 2005b) using a complementary approach.\footnote{
   Note that Rupke \et 2005b quote SFR's for the mass range 0.1 to 100\msun\
   but model energy and momentum injection using Starburst~99 models, which normalize
   the SFR to the mass range from 1 to 100\msun.  These SFR scales differ by a factor
   of 2.55 complicating comparisons to the work presented here.}
This conclusion is preliminary for two reasons however. First, the wind 
efficiency in the hot and warm phases is not 
empirically established for ultraluminous galaxies. Second, the mass of the cool 
outflows could be greatly underestimated if some unresolved regions of the 
absorption troughs are highly saturated.  While the data are largely
consistent with supernova feedback accelerating the cool outflows via the
ram pressure of a hot wind, several alternative acceleration mechanisms
are not yet ruled out.  In particular, the viability of merger-induced winds
is explored in \S~\ref{sec:merger} in light of the rotation observed across
several outflows.

\subsection{Supernova-Driven Winds or Merger-Induced Winds?} \label{sec:merger}

Recent theoretical work suggests outflows might be driven directly by galaxy 
mergers (Cox \et 2004). The collision of gas-rich galaxies may generate
strong shocks, which heat the interstellar gas at the expense of orbital 
energy.  The high pressure would drive an outflow in much the same manner as 
the hot interior of a superbubble does. These merger-driven winds are not easy 
to distinguish from supernova-driven winds because the shock velocities are 
$few \times 100$\kms\ in both cases, generating hot \x emitting gas at $10^6$ 
to $10^7$~K.  

The primary distinguishing factors are (1) the spatial scale of the
energy-injection region and (2) the relation between hot gas mass 
and the star formation rate.  
The size of the energy injection region
will be larger for merger-driven winds than for winds driven by
a nuclear starburst.  The latter develop a bipolar shape because 
the higher density of the disk stalls the shock in the disk
plane (de Young \& Heckman 1994).  If the energy source, however,
is comparable to the size of the disk, the outflow is more
spherical (Cox \et 2005).  The morphology of X-ray surface
brightness contours may eventually  help distinguish these geometries.
The high fraction of ULIGs presenting outflows in \nad, 
16 of 18 objects in the CO-subsample and a similar fraction in
Rupke \et (2005b), requires a wide opening angle for the outflows; 
but the large angular extent of the blueshifted absorption regions 
indicates a cylindrical or spherical, rather than conical, outflow geometry. 
The rotation detected in a few cool outflows, \i1720-00 and \i0015+54,
would also be consistent with a wind geometry similar to a very thick disk.
These geometrical constaints could well reflect an extended, global
starburst rather than a merger-driven wind.
If merger-driven winds are important at all, energy arguments suggest
it is during mergers of  massive, gas-rich galaxies like ULIGs.
Cox \et (2004) found that the temperature and amount of shocked gas 
varied strongly with the type of orbit and impact parameter, but
the total gas consumption was only weakly correlated with the orbit.
Hence, the properties of the hot wind will vary greatly among ULIGs
with similar SFRs if the bulk of the heating is due to merger shocks.

The measured velocities and  masses  were combined to describe the 
kinetic energy of the bulk flow as $KE \approx 0.5 \sum M_i v_i^2$, where the sum was 
over the circular annuli. This {\it observed} kinetic energy is strictly a lower limit since 
the velocities are only line-of-sight components.  Table~\ref{tab:nad} summarizes the results.
For  \i1056+24 and \i1720-00, for example, the measured kinetic energy is $\sim 5 \times 10^{55}$~ergs
and  $\sim 4 \times 10^{55}$~ergs, respectively. For comparison, a Starburst~99 (Leitherer \et 1999) 
population synthesis model (Salpeter IMF and solar metallicity) normalized to the star formation 
rate of \i1056+24 and \i1720-00 indicates that over a period of 10~Myr, supernovae and stellarwinds
supply a mechanical energy of $3.0 \times 10^{57}$~ergs and $7.6 \times 10^{57}$~ergs. Figure~\ref{fig:ke}
confirms that cool outflows typically carry a few percent of the supernova energy.
The starburst phase may last longer, perhaps 100~Myr as favored by some dynamical models 
(Murphy \et 1996), increasing the available feedback energy and   proportionately reducing the inferred efficiency
of converting it into bulk motion of the cool wind.  No evidence for 
a higher thermalization efficiency in ULIGs relative to dwarf starbursts, as suggested 
by Rupke \et (2005b),  is seen in Figure~\ref{fig:ke}. This discrepancy is addressed further in \S5.2.1 
as the two studies agree on the relevant conlusion for this section. 
Adiabatic, wind-driven bubble models can transfer close to 20\% of the supernova energy to the kinetic 
energy of the swept-up shell (Weaver \et 1977); hence starbursts are more than capable of 
powering the observed outflows.  Although the orbital energy of the merger is larger than the
measured kinetic energy of these cool outflows, the thermalization efficiency of merger-driven winds
has yet to be established. Considering that the supernova-driven wind model successfully explains many
properties of the warm and hot outflows observed in starburst galaxies, it remain the favored
power source for the cool outflows, although the momentum supplied may be more important than
the energy as described below.


%

\subsection{Gas Dynamics of Cool Outflows}  \label{sec:dynamical}

How far galactic outlows travel is of great interest. In addition
to regulating the galactic gas mass, galactic winds are thought
to pollute the intergalactic medium with heavy elements.  In Paper~I,
the terminal velocities of the cool outflow were shown to be
near the escape speed. On the one hand, little gas is detected
that is moving fast enough to escape the halo potential. On the
other hand, once  material is moving fast enough to leave the galaxy, 
it will not be further accelerated. Understanding the fate of the
outflowing material therefore requires a dynamical model. Empirical
data for cool outflows should help us identify the best model. 

Murray \et (2005) show that measurements of mass loss efficiency 
would, in principle, distinguish winds driven by momentum deposition
from energy-driven winds.  If a fraction $\xi $ of the mechanical 
energy supplied by supernovae and stellar winds is thermalized, then 
the maximum mass loss rate in an energy conserving wind is
$0.5 \dot{M}_w v_{\infty}^2 = \xi L_w$, where $v_{\infty}$ is the
terminal velocity of the wind. Taking $\xi = 10\%$, 
the mass flux in an energy-conserving wind relative to the SFR is
\begin{eqnarray}
\frac{\dot{M}_w}{\dot{M}_*} = 0.985 \xi_{0.1} \left( \frac{300 {\rm ~km~s}^{-1}}{v_{\infty}} \right)^2,
\end{eqnarray}
where the mechanical luminosity was taken from a population synthesis 
model with  continuous star formation for a period of 10~Myr, a Salpeter initial
mass function, and upper/lower mass limits of 1\msun\ and 100\msun (Leitherer \et 1999). 
The coefficient would be 2.7 times larger at an age of 100~Myr. The cooling
time of the gas associated with the \ion{Na}{1} absorption is likely
shorter than the dynamical timescale of the outflow.   

Cool clouds of gas and
dust are thought to be accelerated by the ram pressure of the hot wind and/or radiation pressure.
For a total momentum deposition rate $\dot{p}_{SB}$, the maximum mass
flux is $\dot{M}_c$, where $\dot{M}_c v_{\infty}  = \epsilon \dot{p}_{SB}$.  
Using the mass flux and energy deposition rates from Starburst~99
(Leitherer \et 1999), the estimated momentum deposition rate is
$\dot{p}_{SB} = \sqrt{2 L_w \dot{M}}$, which is  2.1~$\times 10^{33}$~dyne
at an age of 10, or about 2.4 times higher at an age of 100~Myr.
Since some of the supernova and stellar wind momentum vectors will cancel out in the 
starburst region, this estimate is an upper limit on momentum production. 
For a burst duration of 10~Myr, the 
normalized mass outflow rate of the momentum-conserving solution  is
\begin{eqnarray}
\frac{\dot{M_c}}{\dot{M}_*} = 1.1 \epsilon \left( \frac{300 {\rm ~km~s}^{-1}}{v_{\infty}}
\right).
\end{eqnarray}
In principle, it should be possible to determine whether the cool 
wind is momentum-driven by measuring the ratio of the mass loss rate to the SFR,
i.e. the mass loss efficiency $\eta$, as a function of escape velocity (or
terminal velocity).  These energy-driven and momentum-driven wind models are 
illustrated for reference in Figure~\ref{fig:mm} for comparison to the data.

\subsubsection{Outflow Rates in Cool Phase}

The mass-loss rate is $\dot{M_c}(r) = \Omega \mu m_H n_H(r) v r^2$
for a constant velocity, radial outflow into solid angle $\Omega$.
For a mass flux independent
of radius, the column density through the wind is
\begin{eqnarray}
N_H = \left( \frac{\dot{M} }{ \Omega \mu m_H v} \right)
\left( \frac{r_{max} - r_0}{r_0 r_{max}}  \right).
\end{eqnarray}
Assuming a spherical wind, and taking $\mu = 1.4$ to include the helium mass, the
mass flux in the cool wind is
\begin{eqnarray}
\dot{M_c} = 141 \msunyr \left( \frac{N(H) }{ 4.9 \times 10^{20} 
{\rm ~cm}^{-2}}  \right) \left( \frac{ r_0 }{  5 {\rm ~kpc}} \right) \times 
\nonumber \\
\left( \frac{v }{ 400\kms}  \right)
\left( \frac{\Omega }{ 4\pi } \right).
\label{eqn:mdot}  \end{eqnarray}
Column~7 of Table~\ref{tab:nad} shows the mass loss rates obtained using the
column densities from col.~6 of Table~3 and the size of the starburst region
from col.~3 of Table~\ref{tab:ha}. The mass loss rates (in the cool phase)
reach 10\% of the star formation rate in many ULIGs.\footnote{This statement assumes 
  a lower mass cut-off of 1\msun\ for the initial stellar mass function and
  $dln(N)/dln(M) = -2.35$.} 
The mass obtained by integrating this wind density
over an outflow of opening angle $\Omega$ from a launch radius $r_0$ to $r_{max}$ is

\begin{eqnarray}
M_c =  6.89 \times 10^7\msun\ \left( \frac{r_0}{ 1~{\rm kpc}} \right) 
\left( \frac{r_{max} - r_0}{1 ~{\rm kpc}} \right)
\left( \frac{\Omega}{4\pi}   \right) 
\times \nonumber \\
\left( \frac{N(H)}{4.9 \times 10^{20} {\rm ~cm}^{-2}}  \right).
\label{eqn:mass} \end{eqnarray}
Taking the starburst radius from col.~3 of Table~\ref{tab:ha}
and the maximum radius from col.~4 of Table~\ref{tab:nad},
which is a lower limit on the extent, yields the mass estimates
in col.~8 of Table~\ref{tab:nad}.  The results are similar to
the masses obtained by direct integration of the column density.


These mass loss efficiencies are normalized by the star formation rate
(of 1 to 100\msun\ stars) and plotted with the analogous values
for the  Schwartz \& Martin (2004) galaxies in Figure~\ref{fig:mm}. The
terminal velocities, i.e. the most blueshifted part of the absorption trough,
are much lower in the lower mass galaxies  than in the ULIGs.  Both sets of 
data are consistent with the tracks for momentum-driven winds with
$\epsilon \sim 10$ to 100\%, but the lower efficiencies appear to be
favored in the  ULIGs.  Non-unity values for $\epsilon$ could reflect cancellation 
of the momentum vectors from multiple supernovae in the starburst 
region or that the hot phase of the wind carries more momentum.
The variation of $\epsilon$ with mass is not well-determined. In addition to
uncertainties about the sizes of the absorbing shells in dwarfs, the ionization/depletion 
corrections are highly uncertain. In particular, Schwartz \& Martin adopted $f_d f_{ion} =
100$, but a value of 10 is used here for the ULIGs.  These assumptions yield, for 
purposes of illustration, a higher mass-loss efficiency, $\eta = \dot{M_c}/\dot{M_*}$, 
in the dwarf starbursts than for the ULIGs in Figure~\ref{fig:mm}.  Constant $f_d f_{ion}$, 
in contrast, indicates similar  momentum deposition efficiency, $\epsilon  \sim 10\%$, in 
all the starburst galaxies and reduces the variation  in mass loss efficiency. 

Rupke \et (2005b) agree with these results for the efficiencies of momentum, energy, and mass 
deposition in the cool phase of outflows (in the order of magnitude sense or better) . 
However, their paper also claims more efficient energy deposition in larger starbursts.
In Figure~\ref{fig:mm}, the higher value of $f_d f_{ion} = 100$ for the dwarf galaxies
yields constant thermalization efficiencies $\xi \sim 1\%$. Lowering the ionization
and depletion corrections in the dwarf starbursts (to match the ULIG values) qualitatively 
reproduces the trend noted by Rupke \et (2005b).  The scalings cannot be compared
quantitatively because Rupke \et do not consistently use the same lower mass cut-off
for the ULIG SFR's, the dwarf galaxy SFR's, and the star formation rate of the population
synthesis model..  The robust results from
Figures \ref{fig:mdyn}, \ref{fig:ke}, and ~\ref{fig:mm} and Rupke \et (2005b) are
the efficiencies of mass, energy, and momentum deposition in the {\it cool phase} of
ULIG outflow.  More measurements are needed for the dwarf starbursts, although it
appears at least plausible that the efficiencies do not vary much.



\subsubsection{Turn Around Radii}

%

To explore how far the cool clouds in ULIG outflows might travel, 
the momentum-driven-wind models described by Murray \et (2005) were
used to estimate turnaround radii. The momentum deposition falls off 
as $r^{-2}$ for both ram pressure driven winds and winds driven by 
radiation pressure. Maximum acceleration takes place near the base of 
the free-flowing wind, and the gravitational potential of a roughly
isothermal halo will decelerate the flow at large radii. 

Consider the motion in an isothermal halo of circular velocity 
$V_c = 250$\kms, or velocity dispersion $\sigma = 177$\kms. 
For ram pressure driven winds,  the clouds begin to decelerate at a radius,
\begin{eqnarray}
R_g = 37 {\rm ~kpc} \frac{\dot{M}_*}{100 \msunyr} \left(
\frac{177 {\rm ~km~s}^{-1}}{\sigma} \right)^2 \times \\
\frac{2.5 \times 10^{20}\col}{N_H}. 
\end{eqnarray}
Integrating the equation of motion from a launch radius of 1~kpc,  the 
turnaround radius is found to be so large that the clouds 
are not likely to fall back. For radiation pressure driven winds in 
the optically-thin limit, the deceleration radius is
\begin{eqnarray}
R_g = 3.6 {\rm ~kpc} \frac{{L}}{10^{12}\lsun} \left(
\frac{177 {\rm ~km~s}^{-1}}{\sigma} \right)^2 \times \\
\frac{\kappa}{650 {\rm ~cm}^2 {\rm ~g}^{-1}}. 
\end{eqnarray}
Integrating the equation of motion, one finds the turnaround radius is 
about 32~kpc. 

It is not currently known which of the above wind models, or an
optically thick radiatively-driven wind model, is a better description of cool 
outflows in ULIGs. However, they all have large enough turn-around radii to 
present a significant geometrical cross section for absorption. To illustrate
the space density of winds, consider a fiducial turn-around radius of 30~kpc 
for the cool outflow, which is a conservative, lower limit based on the 
dynamical model above.


\subsubsection{Intervening Absorption from Cool Winds}

The association of intervening absorbers with galaxies is well-established
for strong, low-ionization (e.g. Mg~II) systems. The open question is whether 
these absorbers are accreting halo gas or starburst winds. The properties of 
low-ionization gas in outflows was not easy to predict, however, due to the
non-equilibrium ionization state in winds and the multi-phase fluid.  Although
the observations presented in this paper are taken along sightlines with a special
orientation, they do provide some constraints.

In the local universe, the space density of infrared bright galaxies exceeds 
that of other classes of galaxies for $L_{Bol} > 10^{11}$\lsun\ (Soifer \et 
1986; Soifer \et 1987). Adjusting the Soifer \et  luminosity function to 
$H_0=70$\kms Mpc$^{-1}$, the local number density of galaxies brighter than
$10^{11}$\lsun\ is $1.0 \times 10^{-5}$~Mpc$^{-3}$.  For an absorber
radius of 30~kpc, the redshift path density is small,
$dN/dz \approx 1.4 \times 10^{-4}$, and approximately equal
to $n_0 \sigma_{cross} c H_0^{-1}$ (within 15\% at z=0.1).
If the infrared active phase is, however, a nonrecurring event, the 
fraction of normal galaxies passing through this stage approaches
unity for infrared-bright periods lasting less than 100~Myr
(Soifer \et 1987).  For cool ULIG winds to produce a significant 
number of intervening absorbers,  the winds must hang around the
galaxy long after the starburst ends.  

These relics should be observable as intervening MgII systems. Steidel
\& Sargent (1992) found the redshift density of strong, $W_r > 1.0$\AA,
Mg~II systems was $dN/dz = 0.27$ over $0.2 < z < 2.2$. A source density 
of $ n_0 = 0.022 {\rm ~Mpc}^{-3} (30~{\rm kpc}/r)^2$ is required to
create a comparable path density. For the space density of normal
galaxies to be this high (Blanton \et 2003), galaxies as faint as
$0.011 L^*$ would need to contribute to the relict wind population.
This estimate can be improved by applying the measured evolutionary 
dependence of strong MgII systems (Prochter \et 2004). Taking into account
the increase in path density with redshift, the extrapolation to 
redshift 0.1 reduces the number density of sources to 
$ n_0 = 0.0087 {\rm ~Mpc}^{-3} (30~{\rm kpc}/r)^2$, which is 
similar to the density of galaxies brighter than $0.12L^*$.  
In other words, if every $L > 0.1 L^*$ galaxy goes through a 
starburst phase, and remnants of the  cool outflows persist into
the post-starburst phase, then relict winds could account for
the entire path density of strong Mg~II systems at low redshift.
Larger cross sections clearly raise the minimum mass of galaxies that
would need to contribute relic outflows to the path density. 

This hypothesis needs to be tested further against two additional
constraints. First, the local path density of OVI systems, which 
may trace the shocks from these winds, also suggests dwarf galaxies 
make a large contribution to the \ion{O}{6} cross section 
(Danforth \& Shull 2005; Stocke \et 2005). However, to date, direct 
absorber -- galaxy associations require cross 
sections $\sigma_r \sim 200$~kpc (Adelberger \et 2005; Stocke 
\et 2005;  Tumlinson \& Fang 2005), which is plausible but not
yet required for the cool outflows. Second, and more importantly,
the velocity widths of low-ionization, intervening systems need
to be compared to the distribution measured for outflows, which
has a mean here of 350\kms\ FWHM.  Songaila \et (2005b) 
suggest that the linewidths and ionization state of absorbers
can differentiate winds from other types of halo gas.




%




\section{Conclusions}

This work maps the outflow properties across 18 ultraluminous infrared-selected 
galaxies, providing measurements of \nad\  column density, 
\nad\ absorption-line velocities, and \Ha\ emission-line velocities.  These starburst 
galaxies are undergoing major mergers, as evidenced by the presence of multiple nuclei 
and prominent tidal features. The cool outflows previously identified in absorption against 
the stellar continuum (Heckman \et 2000; Rupke \et 2002;  Martin 2005; Rupke \et 2005a, 2005b) 
are shown to cover a large portion of the galaxy, exceeding  15~kpc along the slit in a few cases.
Faint, spatially-extended \Ha\ emission, often diffuse emission or narrow spikes in the 
spectra rather than well-defined Doppler ellipses as found in \i0315+42, is detected in 
about half of the ULIGs;  but its presence shows no obvious correlation with the strength 
of the \nad\ absorption. The gas velocity in the blue wing of the \Ha\ line profile is 
comparable to,  and may exceed (Rupke \et 2005b),
that of the most blue-shifted absorbing gas. It follows that 
the outflow probably does contain warm \Ha-emitting gas; but the emission line profile is 
largely shaped by higher density gas.

The absorption profiles extend from the systemic velocity up to (blueshifted) velocities 
comparable to the halo escape velocity, suggesting acceleration of cool, interstellar gas.  
Across many ULIGs, \i1056+24 being one example, the outflow velocity is constant over the 
entire galaxy while the \Ha velocity shows a marked rotational gradient, shifting from 
negative to positive velocities across the nucleus.  These results are consistent with an 
outflow diverging from a central starburst, because the rotational velocity of a diverging 
wind slows down with increasing height to conserve angular momentum.  Surprisingly, however, 
across several ULIGs, particularly \i1720-00 and \i0015+54, the cool outflows present a 
velocity shear as large as the projected galactic and orbital rotation combined. These
marked kinematics suggest a low scale height and large launch region for at least
these outflows.  It will be interesting to see whether numerical simulations of winds
from galaxy mergers can reproduce such results by including spatially extended star formation
or merger-driven outflows. If not, coeval winds from both nuclei might offer the only
explanation as overlapping disks do not work in general.

%
%

The \nad\ $\lambda \lambda 5890, 5896$ lines are blended in ULIG spectra, due to 
the presence of multiple absorption components along the sightline.  Where this
absorption trough was highly structured, the doublet ratio yielded optical depths
$\sim 2$ to 10. Less structured troughs allowed much larger optical depths, which is
problematic due to the nonlinear relation between the doublet ratio and the 
column density.  This degeneracy was resolved by fitting the line optical 
depth, rather than the doublet ratio.   The fractional area of the beam 
covered by absorbing gas was estimated from the residual line intensity before
calculating the optical depth profiles from the intensity profiles. This approach
produced column densities, $N(\nad)$,  between  $ 1.8 \times 10^{13}$~cm$^{-2}$ and
$2.4 \times 10^{14}$~cm$^{-2}$.  A few systems common to the Rupke \et 
(2005a) sample, and our \nad\ column density measurements agree to within a factor of 
two.

The mass of outflowing gas, estimated by integrating column densities 
over the area of the outflow, is of order $10^8$\msun, or a few percent
of the gas mass, in most ULIGs.  These values are lower limits in several systems
where the projected outflow appears to be larger than the background continuum 
source.  The reddening increases with the strength of the \nad\ 
absorption as was found previously in LIGs (Heckman \et 2000) but exceeds 
the amount expected for the estimated outflow columns and a Galactic dust-to-gas 
ratio.  The applied corrections for Na ionization and depletion could be low,
indicating an even higher gas column, but this is not necessary as some of the
dust may not make it into the wind, just as much of the  interstellar gas is
left behind.  Previous empirical results suggested mass outflow rates in 
the {\it hot and warm} phases were comparable to the SFR  in starburst
galaxies (Martin 1999), although this is not well-constrained in galaxies
as luminous as ULIGs.  Based on a  simple model for a freely-expanding outflow,
the measured gas columns and velocities indicate mass loss rates in the cool
gas of order 10\% the SFR, in rough agreement with  Rupke \et (2005b).
This efficiency would be higher in dwarf starbursts than ULIGs 
if ionization and depletion corrections are actually larger in the former as
suggested. Such uncertainties are currently too large to allow a definitive
discrimination between momentum-driven and energy-driven outflows in the cool
gas.

The fate of the cool outflows remains model dependent for now. Merger-driven winds 
appear unlikely to eject a lot of material from the halo potential, and their viability is not 
yet certain. It is clear that momentum-driven winds would easily transport the gas detected 
in \nad\ absorption to radii $\sim 30$~kpc, and much further for some choices of parameters. 
The momentum carried by the cool winds is roughly 10\% of that supplied by supernovae and
stellar winds (or the radiation field); and  the kinetic energy carried by the cool gas 
is a few percent of the supernova energy  (for a burst age of 10~Myr). These measured
efficiencies provide a guide for feedback recipes in galaxy formation models, which
should distinguish the cool and hot gas. They appear consistent with  estimates for 
LIGs and ULIGs presented by Rupke \et (2005b) and theoretical expectations for
supernova-driven winds.

The interpretation of the data presented here could be improved by 
new numerical simulations. Numericists are beginning to address
the temporal evolution of outflows (Hopkins \et 2005), 
spatially extended star formation (Fragile \et 2004),
and merger-driven winds (Cox \et 2004; Cox \et 2005). It
will be interesting to learn what rotational properties these
models generate for the entrained gas component, and whether
clouds dense enough to produce \nad\ absorption survive in the
wind.  More detailed simulations of merger-driven winds are needed 
to determine their viability, particularly quantitative
predictions for the efficiency with which orbital kinetic energy 
is transferred to the bulk motion of the outflow.  More
observations of thermal \x emission from ULIGs will distinguish whether a hot
wind plays an important role in the acceleration of the cool gas, as the hot
phase is not required in a purely radiative model of acceleration.

The measurements presented here suggest the cross sections and column densities for the 
cool component of winds are significant. 
Ultraluminous infrared galaxies are rare in the local universe, but such extremely
luminous starbursts were much more common in the past.  If most galaxies brighter
than 0.1~$L^*$ have undergone such extreme star formation activity, the relic winds
should be detectable as intervening absorption-line systems.

\acknowledgements{The author thanks TJ Cox, Tim Heckman, 
Norm Murray, Todd Thompson, Eliot Quataert, and Sylvain Veilleux
for stimulating discussions.
The R band images shown in this paper were kindly provided by
Lee Armus and Tom Murphy. The author thanks Dr. Colleen Schwartz for flux
calibrating the spectra. This paper was written largely while the author was 
a visitor at the KITP. This work was supported by grants from the David and 
Lucile Packard Foundation and the Alfred P. Sloan Foudation.}



\appendix

\begin{center}
{A SIMPLE TOY MODEL FOR THE ROTATION OF CONICAL WINDS}
\end{center}

%

When clouds launched from a galactic disk follow ballistic trajectories 
through the halo, they rotate more slowly than the gas disk (Bregman 1980; 
Collins \et 2002).  A starburst-driven wind differs from these fountain 
models in that the starburst outflow is launched from a much smaller
region of the disk. The expansion of the superbubble quickly stalls in the 
dense gas disk, and the shell develops a bipolar shape (DeYoung
\& Heckman 1994; Strickland \& Stevens 2000; MacLow \& Ferrara 1999;
Mori \et 2002, Fragile \et 2004).  After the elongated supershell erupts from 
the galaxy, the outflow is mostly radial (Chevalier \& Clegg 1985). 
Rotation of the outflow has been ignored in these calculations 
largely because it is expected to be small.  

To provide a baseline model for the rotational properties of galactic
outflows, as measured by absorption-line position-velocity diagrams,
this Appendix considers the rotation of galactic winds in two limits:
(1) constant angular momentum, and (2) constant angular velocity.  
Angular momentum conservation provides the simplest model. Outflows 
with constant angular velocity provide some insight into the rotational 
properties of magnetically-driven winds. Only when such torques act on the 
outflow, will significant rotation be observed in projection along the
major axis of the disk. When the velocities are projected against the minor 
axis, a velocity jump can be seen across the nucleus, in the absence of any
torques. The argument goes as follows.

\section{A1. OUTFLOWS CONSERVING ANGULAR MOMENTUM} 

Consider the trajectory of clouds launched from a central starburst in 
a disk, which rotates at speed $V_c$.  Figure~\ref{fig:toymodel} sketches 
the geometry.  The starburst region is described 
by a squat cylinder of radius $R_{0}$. Disk gas entrained in the wind
at $R_0$ flows freely outward at speed $V_w$, and this outflow diverges
radially along a conical surface of opening angle $\theta_w$. The
thickness of the starburst region is $z_{0} = R_{0} / tan~ \theta_W$. 
As a ring of wind material moves outward along the cone, the circular 
velocity of the gas declines to conserve angular momentum as $V(R) = 
V_c R_0 / R$, where $R$ is the radius in cylindrical polar coordinates. 

Now consider what an observer viewing this simple toy model would
see. The location of the observer in \fig~\ref{fig:toymodel} defines 
the inclination of the galaxy, $i$, which is the angle between
the sightlines and the polar axis of the galaxy. A slit aligned
with the major axis of this galaxy samples the model along a
series of parallel sightlines that intersect the x-axis (see
\fig~\ref{fig:toymodel}). The intersection of these sightlines
with the surface of the conical outflow define the locus 
where the velocity field is sampled. In each plane, $x=constant$,
the intersection of the sightline and the hyperbolic conic section
is easily computed.  At these points, the projection of the circular
velocity along the sightline is
\begin{eqnarray}
V_{c,proj} = V(R) cos (\phi) sin(\iota),
\end{eqnarray}
and the projection of the outflow velocity is
\begin{eqnarray}
V_{w,proj} = V_w sin(\theta_w) sin(\phi) sin(\iota) + V_w cos(\theta_w)
cos(\iota).
\end{eqnarray}
For a wind half-opening angle of $\theta_w = 60\deg$, the resulting 
position-velocity diagram for projection along the major
axis is shown in \fig~\ref{fig:vx3}a.
The rotation speed of the disk changes sign on the approaching 
and receding sides of the major axis, but the projected speed of the radial 
wind is symmetric to either side of the minor axis.  Along the major 
axis of a galaxy, the outflow velocities measured in absorption will not 
present much of a velocity gradient compared to the rotation curve of the 
disk provided $V_w \ge\ V_c$. 


For sightlines intersecting the minor axis,  the rotational velocity 
is perpendicular to the sightline and therefore undetectable. Projection
of the radial outflow gives observed velocities 
\begin{eqnarray}
V_{w,proj} = V_w (sin(\theta_w) sin(\iota) +  cos(\theta_w) cos(\iota)), y \ge\ 0\\
=V_w(-sin(\theta_w) sin(\iota) + cos(\theta_w) cos(\iota)), y < 0,
\end{eqnarray}
along the minor axis (y-axis in \fig~\ref{fig:toymodel}). The P-V
diagram along the minor axis  is not symmetric about the nucleus. 
On the near side of the 
minor axis, where the cone is tipped toward us, the apparent velocity may 
approach $V_w$. In contrast, on the far side, the radial velocity vector 
makes a larger angle with the sightline, so the projected speed is lower.
Figure~\ref{fig:vx3}b illustrates the jump in the projected 
wind speed along the minor axis for several viewing angles. Notice that the 
wind can appear redshifted against the far side of the disk when the viewing 
angle $i$ is close to the opening angle of the cone -- 
i.e. $90\deg - \theta_w < i < \theta_w$. In summary, although the wind 
velocity varies little along the major axis, it can jump abruptly along the 
minor axis near the nucleus. The P-V diagram would not show a 
{\it smooth} velocity gradient along the entire minor axis.


\section{A2. CONSTANT ANGULAR VELOCITY WIND}


This physically-motivated, but purely geometrical, model provides some 
intuitive feel for how projection affects the observed wind velocities.
Combined with observations, it suggests some outflows acquire additional 
angular momentum as they leave the disk.
Magnetic field lines anchored in the disk but rising into the outflow cone
would act somewhat like rigid wires because the gas is forced to corotate 
with the field lines out to the Alfv\'{e}n point (Pino \& Tanco 1999). 
Magnetic fields might therefore spin up the outflow at the expense of the 
disk angular momentum.  To illustrate the potential impact on the observed 
velocity gradient, the toy model was given a constant angular velocity. 
The circular velocity at every point  was  set to $V_{\phi} = V_C 
R / R_0$.  The wind velocity is again radial and was taken to be 
approximately constant at $V_w$, ignoring magnetic slingshot effects.
The blue lines in \fig~\ref{fig:vx3}a illustrate the projected P-V diagram 
along the major axis.  The projected velocity gradient is much steeper  
than the constant-L model.  For a wind speed twice that of the disk rotation,
the outflow appears blueshifted, toward the receding side of the major
axis, out to two starburst radii.  {\it This example demonstrates that 
outflows with constant angular velocity can produce blueshifted absorption 
with large velocity gradients.} The presence of halo magnetic fields is 
supported by observations of some starbursts, which show extended 
synchrotron emission (e.g. Thompson \et 2006; Israel \& deBruyn 1988; Dahlem \et 1995).




\clearpage


\newpage


\begin{table}
\caption{Properties of Extended \nad\ Absorption}
\begin{tabular}{llclllcl}
\hline
 Galaxy    &       Slit PA  &   Size   & Size  & $M_{Dop}$ & KE     &  $\dot{M}_{Dop}$    & $M_c$\\
           &       (\deg)   &   (\asec)    & (kpc)     & (\msun)   & (ergs) & (\msunyr)           & (\msun) \\
  (1)      &  (2)           &   (3)        & (4)       & (5)       & (6)    & (7)                 & (8)  \\
\hline
\hline
IRAS00153+5454 & -22.0    & 7\farcs3\tablenotemark{a} & 15             & $1.8\times 10^{8}$                 & $1.6 \times 10^{55}$  & 2          & $2.5\times 10^8$ \\
IRAS00188-0856 & -3.3     & 1\farcs9                  & 4.4             & $2.9\times 10^{8}$                  & $3.2 \times 10^{56}$  & 5          & $5.6\times 10^7$ \\
IRAS00262+4251 & -14.1    & 3\farcs8                  & 6.8             &$2.7\times 10^{7}$                   & $1.6 \times 10^{55}$  & 11         & $3.4\times 10^8$ \\
IRAS03158+4227 & -15.0    & 2\farcs6                  & 6.2             &$4.7\times 10^{7}$                   & $5.6 \times 10^{55}$  & 11         & $1.2\times 10^8$ \\
IRAS03521+0028 & 76.5     & 1\farcs9                  & 5.0             &   \nodata                           & \nodata               & \nodata    & \nodata          \\
IRAS08030+5243 &  0       & $\sgreat 5\farcs2$        & $\sgreat 8.2$   &$\sgreat 1.6\times 10^{7}$           & \nodata               & $ 3$       & $7.7\times 10^7$ \\
IRAS10494+4424 &  25.0    & $\sgreat 7\farcs0$        & $\sgreat 12$  &$\sgreat 5.9\times 10^{7}$           & $1.4 \times 10^{55}$  & $\sgreat 3$& $1.2\times 10^8$ \\
IRAS10565+2448 &  109.0   & $\sgreat 13\farcs4\tablenotemark{a}$ & $\sgreat 11$  &$\sgreat 1.1\times 10^{8}$&  $4.6 \times 10^{55}$ & $\sgreat 8$& $2.3\times 10^8$ \\
IRAS11506+1331 &  80.3    & 5\farcs9\tablenotemark{a} & 13            &$1.74\times 10^{8}$                  & $4.3 \times 10^{54}$  & 0.72       & $3.2\times 10^7$ \\
IRAS15245+1019 &  127.8   & $\sgreat 12\farcs4$       & $\sgreat 17.8$   & $\sgreat 6.7\times 10^{7}$         & $2.8 \times 10^{55}$  & $\sgreat 6$& $4.4\times 10^8$ \\
IRAS16090-0139 &  107.6   & 3\farcs2                  & 7.6             & $8.0\times 10^{7}$                  &  $7.0 \times 10^{55}$ & 8          & $1.8\times 10^8$ \\
IRAS16487+5447 &   66.2   & 5\farcs7                  & 11            &  \nodata                            &  \nodata              & \nodata    & \nodata          \\
IRAS17208-0014 &  166.7   & $\sgreat 8\farcs6$                  & $\sgreat 7.3$   & $7.0 \times 10^{7}$                 & $3.8 \times 10^{55}$  & 7          & $1.2\times 10^8$ \\
IRAS18368+3549 &  -31.0   & 4\farcs8                  & 10            & $1.4\times 10^{8}$                  &$1.2 \times 10^{56}$   & 5          & $1.9\times 10^8$ \\
IRAS19297-0406 &  149.5   & 2\farcs9                  & 4.8             & $4.0\times 10^{7}$                  &$5.8 \times 10^{55}$   & 5          & $5.3\times 10^7$ \\
IRAS19458+0944 &  118.1   & \nodata                   & \nodata         & \nodata                             & \nodata               & \nodata    & \nodata        \\
IRAS20087-0308 &   85.5   & $\sgreat 6\farcs3$        & $\sgreat 12$  &$\sgreat 9.5\times 10^{7}$           & $1.1 \times 10^{56}$  &$\sgreat 14$& $3.8\times 10^8$ \\
IRAS23365+3604 &  -19.5   & 4\farcs7                  & 5.9             &$4.4\times 10^{7}$                   & $5.6 \times 10^{55}$  & 8          & $1.1\times 10^8$ \\
\hline
\end{tabular}
\tablecomments{Col 1 - Galaxy name; see Table~1 of Paper~I for redshift and star formation
rate; 
Col 2 - Slit position angle (east of north); 
Col 3 - Angular length of slit where \nad\ absorption detected. Lower limits indicate detections
extend as far as continuum detection.  
Col 4 - Projected spatial extent of the \nad\ absorbing gas, scales as$h_{70}^{-1}$;
Col 5 - Mass of gas detected in absorption.  The corrections for dust depletion and ionization 
correction are as given in Eqn.~\ref{eqn:nh}, and the \nad\ columns from Table~3 column~6 
are adopted.  The area for each aperture is assumed to be an annulus centered on the
nucleus, and the masses of all apertures are summed.
Col 6 -  Kinetic energy estimated from the summed apertures $KE = \sum M_i V_i^2$,
using the same masses as in Col.~6 and the velocities from Col.~9 of Table~3.
Col 7 - Mass loss rate estimated from a free-streaming wind model normalized to the
column density and velocity in the nuclear aperature, as described by Eqn.~\ref{eqn:mdot}.
Col 8 - Mass in wind estimated from the model described by Eqn.~\ref{eqn:mass}. The 
launch radius is taken from Col.~3 Table~\ref{tab:ha} , and the maximum extent is taken from
Col.~4 above.
}
\tablenotetext{a}{Detected \nad\ toward two nuclei, and the entire extent of the absorption is listed.
}
\label{tab:nad} \end{table}


\begin{table}
\caption{Emission Properties}
\begin{tabular}{lllllccl}
\hline
 Object	        &  Scale                & $R_{CO}$     & \Ha Extent    & $\nabla v_{H\alpha}$ (km  & $\nabla v_{NaI}$ (km  &  c(\Hb) & Merger \\
	        & (kpc/\asec)           & (pc)         &(kpc)          & s$^{-1}$ ~kpc$^{-1}$)                  &  s$^{-1}$ ~kpc$^{-1}$)  &         & State \\
	(1)     & (2)                   & (3)          &(4)            & (5)                                & (6)                & (7)     & (8) \\
\hline
\hline
IRAS00153+5454  &     2.038 & \nodata  & 25.5   (blob)   & 7  (17, 12 \& 7)  &  $16\pm4$  &$1.41 +   0.82   $  &  double \\            
IRAS00188-0856  &     2.295 & 297      & 10  (halo)    & 0	         &  $7\pm16$  &$2.42 +   2.42   $  &  double \\                 
IRAS00262+4251  &     1.799 & 314      & 19.8 (halo)   & 15	         &  $7\pm19$  &$0.99 +   0.37   $  &  tails \\                  
IRAS03158+4227  &     2.384 & 405      & 31.6  (spike)    & 16	         &  $40\pm14$  &$1.72 +   3.05   $  &  single \\                 
IRAS03521+0028  &     2.641 & 575      & \nodata		    & \nodata         	 &  \nodata  &$2.81 \pm 2.81$ &  irreg \\
IRAS08030+5243  &     1.570 & 345      & 19.9 (blobs)  & 60 	         & $8\pm3$  &$1.33 +   0.45   $  &  single \\                 
IRAS10494+4424  &     1.717 & 379      & 17.6  (spike)    & 2     &  $24\pm5$  &$1.79 +   3.83   $  &  tails \\            
IRAS10565+2448  &     0.849 & 317      & 6.4 (7\farcs5)   & 13 (11, 19)       &  $9\pm7$  &$1.45 +   0.53   $   & double\\
IRAS11506+1331  &     2.276 & 361      & 27.7  (spike)    & 3 (1, 14 \& 7)    &  $10\pm5$   &$1.23 +  1.30   $   & double  \\         
IRAS15245+1019  &     1.436 & \nodata  & 18.0  (spike)     & 9 (10, 6 \& 21)     & $1 \pm 1$   &$1.47 +  0.69   $   & double  \\        
IRAS16090-0139  &     2.371 & 459      & 20  (spike)     & 8	         &  $19\pm19$  &$0.99 +   2.93   $   & single  \\               
IRAS16487+5447  &     1.903 & \nodata  & 11.6 (diffuse)  & \nodata	         &  \nodata  &$0.41\pm   0.41   $   & double  \\ 
IRAS17208-0014  &     0.846 & 301      & 7.9 	    & 47 (38, 65 \& 20)         &  $42\pm2$  &$1.53 +   2.50   $   & double  \\                
IRAS18368+3549  &     2.104 & 396      & 21.2 (halo)   & 35	         &  $23\pm11$  &$2.6  \pm   2.6    $   & tails  \\               
IRAS19297-0406  &     1.607 & 385      & 15 	    & 24		 &  $2 \pm 9$  &$1.78 +   3.83   $  &  single \\                
IRAS19458+0944  &     1.844 & 465      & \nodata		    & \nodata		 &  \nodata  &$1.79 \pm 1.79$ & tails  \\
IRAS20087-0308  &     1.936 & 466      & 6.5 	    & 26 (15, 25 \& 5)	 &  $19\pm8$  &$4.0  \pm   4.0    $  & double  \\               
IRAS23365+3604  &     1.239 & 372      & 15.4 (blobs)  & 34 (0, 34)         &  $6\pm29$  &$1.20 +   0.50   $  & tails  \\               
\hline
\end{tabular}

\tablecomments{
(1) Object;
(2)  Angular scale at redshift given in Paper I (h = 0.7, $\Omega_0 = 0.3$, $\Omega_{\Lambda} = 0.7$);
(3)  Radius of CO emission from Solomon \et (1997);
(4) Spatial extent of \Ha emission along slit (PA given in col.~2 of Table~\ref{tab:nad}, and slit center 
shown in Figure 1); and description of \Ha\ morphology based on the upper right panels of Figure 1.
(5) Total \Ha  velocity gradient across the ULIG.  For double systems, the projected orbital and 
rotational components are listed, respectively, in parentheses.
(6) Total \nad\ velocity gradient across the ULIG.
(7)  Logarithmic extinction at \Hb  dervied from the measured \Ha/\Hb flux ratio,
an intrinsic ratio of 2.86, and a standard interstellar extinction curve 
(Miller \& Mathews 1972). The difference between this value  and that derived from 
the F(\Hb)/F(\Hg) flux ratio was used to estimate the systematic error in the reddening.
(8) Merger morphology based on  Murphy et al. (2001) scheme and images from Murphy et al. (1996).
}
\label{tab:ha} \end{table}


\clearpage


\begin{deluxetable}{llccclllccllccr}
\rotate
\tabletypesize{\scriptsize} 
\setlength{\tabcolsep}{0.04in} 
\tablewidth{8.7in}
\tablecaption{Na I Spectral Line Fitting}
\tablehead{
\colhead{IRAS} &
\colhead{Ap} &
\colhead{Offset} &
\colhead{$C_f$} &
\colhead{$\tau$} &
\colhead{N(NaI)} &
\colhead{DR} &
\colhead{N(NaI)} &
\colhead{$V$} &
\colhead{$\Delta V$} &
\colhead{min N(NaI)} &
\colhead{max N(NaI)} &
\colhead{$V$} &
\colhead{$\Delta V$} &
\colhead{$V_{H\alpha}$} \\
\colhead{} &
\colhead{} &
\colhead{(\arcsec)} &
\colhead{} &
\colhead{($\lambda5896$)} &
\colhead{(cm$^{-2}$)} &
\colhead{} &
\colhead{(cm$^{-2}$)} &
\colhead{(km s$^{-1}$)} &
\colhead{(km s$^{-1}$)} &
\colhead{(cm$^{-2}$)} &
\colhead{(cm$^{-2}$)} &
\colhead{(km s$^{-1}$)} &
\colhead{(km s$^{-1}$)} &
\colhead{(km s$^{-1}$)} \\
\colhead{(1)} &
\colhead{(2)} &
\colhead{(3)} &
\colhead{(4)} &
\colhead{(5)} &
\colhead{(6)} &
\colhead{(7)} &
\colhead{(8)} &
\colhead{(9)} &
\colhead{(10)} &
\colhead{(11)} &
\colhead{(12)} &
\colhead{(13)}  &
\colhead{(14)} &
\colhead{(15)} 
}
\startdata
0015+54  &  1 &    2.39 & \nodata   & \nodata  & \nodata  & \nodata &        \nodata & \nodata  & \nodata  &  $6.3\times10^{12}$ &  $3.3\times10^{14}$ &  -156$\pm$   191 &  449$\pm$  391 & -150\\
0015+54  &  2 &    0.70 & 0.30   & 6.88  & $7.0\times10^{13}$ & 1.137 &  $4.5\times10^{13}$ &-364 $\pm$       19 &555 $\pm$       54             &  $4.3\times10^{13}$ &  $5.5\times10^{14}$ &  -372$\pm$   21 &   533$\pm$  55 & -143\\
0015+54  &  3* &   -0.98 & 0.35   & 4.79  & $4.9\times10^{13}$ & 1.167 &  $2.6\times10^{13}$ &-276 $\pm$          34 &  497 $\pm$           89 &  $2.1\times10^{13}$ &  $4.2\times10^{14}$ &  -285$\pm$   34 &  462$\pm$  85 & -65\\
0015+54  &  4 &    -4.2 & 0.58   & 4.31  & $4.4\times10^{13}$ &1.177 &  $4.1\times10^{13}$ & -183 $\pm$           20 &  452 $\pm$           48 &  $5.1\times10^{13}$ &  $7.4\times10^{14}$ &   -199$\pm$   22 &  442$\pm$  47 & 28\\
0015+54  &  5 &    -5.9 & 0.40   & 3.12  & $3.2\times10^{13}$ &1.219 &  $3.3\times10^{13}$ & -179 $\pm$           21 &  450 $\pm$           51 &  $1.3\times10^{13}$ &  $3.6\times10^{14}$ &  -158$\pm$   52 &  380$\pm$  110 & -44\\
0018-08  &  1 &    6.99 & \nodata   &  \nodata  & \nodata & \nodata &        \nodata & \nodata  & \nodata  &   $9.1\times10^{9}$ &   $9.1\times10^{9}$ &   315$\pm$   302 &  585$\pm$   1214 & \nodata \\
0018-08  &  2 &    1.64 & 0.24   & 5.55 & $5.7\times10^{13}$ &1.153 &  $1.8\times10^{13}$ & -211 $\pm$          141 &  421 $\pm$          130 &  $7.5\times10^{12}$ &  $2.8\times10^{14}$ &  -214$\pm$   175 &  414$\pm$  140 & -16\\
0018-08  &  3* &    0   & 0.45   & 6.53  & $6.7\times10^{13}$ & 1.141 &  $5.8\times10^{13}$ & -356 $\pm$           13 &  574 $\pm$           19 &  $1.2\times10^{14}$ &  $8.0\times10^{14}$ &  -358$\pm$   13 &  573$\pm$  18 & -33\\
0018-08  &  4 &   -1.72 & 0.17   & 7.33 & $7.5\times10^{13}$ &1.133 &  $2.7\times10^{13}$ & -355 $\pm$           85 &  609 $\pm$          104 &  $1.1\times10^{13}$ &  $3.3\times10^{14}$ &   -355$\pm$  100 &  611$\pm$  109 & -17\\
0026+42  &  1* &   0.48 &  0.14  & 4.24 & $2.4\times10^{14}$  & 1.179 &  $1.3\times10^{13}$ & -211 $\pm$           18 &  387 $\pm$           35 &  $2.1\times10^{13}$ &  $2.4\times10^{14}$ &   -218$\pm$   18 &  384$\pm$  36 & 22\\
0026+42  &  2 &  -1.19 & 0.12   & 4.84 & $1.7\times10^{14}$ &1.163 &  $9.3\times10^{12}$ & -284 $\pm$           21 &  357 $\pm$           37 &  $1.2\times10^{13}$ &  $1.7\times10^{14}$ &  -293$\pm$   21 &  401$\pm$  49 & -53\\
0026+42  &  3 &  -2.88 &  \nodata  & \nodata & \nodata & \nodata &      \nodata & \nodata  & \nodata  &  $9.2\times10^{11}$ &  $4.2\times10^{13}$ &  -263$\pm$   116 &  281$\pm$   124 & -39\\
0315+42  &  1 &    1.68 & \nodata   & \nodata  & \nodata & \nodata &        \nodata &  \nodata & \nodata  &  $3.7\times10^{12}$ &  $6.1\times10^{12}$ &  -708$\pm$   128 &  670$\pm$  221 & 49\\
0315+42  &  2* &    0   & 0.41   &  7.00 & $7.2\times10^{13}$ &1.135 &  $5.7\times10^{13}$ & -526 $\pm$           14 &  461 $\pm$           49 &  $9.6\times10^{12}$ &  $1.0\times10^{13}$ &  -465$\pm$    18 &  514$\pm$  45 & 37\\
0315+42  &  3 &   -1.68 & 0.28   & 7.16 & $7.3\times10^{13}$ &1.135 &  $3.9\times10^{13}$ & -322 $\pm$           28 &  381 $\pm$           61 &  $5.9\times10^{12}$ &  $7.3\times10^{12}$ &  -306$\pm$   51 &  426$\pm$  66 & -27\\
0803+52  &  1 &    1.94 & 0.69   & 3.84 & $3.9\times10^{13}$  &1.194 &  $1.8\times10^{13}$ &  238 $\pm$            4 &  160 $\pm$            6 &  $1.0\times10^{13}$ &  $4.5\times10^{14}$ &   233$\pm$   9 &  164$\pm$   16 & -252\\
0803+52  &  2* &    0   & 0.76   & 4.05 & $4.1\times10^{13}$  &1.187 &  $3.1\times10^{13}$ &  258 $\pm$            3 &  150 $\pm$            4 &  $1.8\times10^{13}$ &  $5.7\times10^{13}$ &   259$\pm$   2 &  147$\pm$  4 & 0\\
0803+52  &  3 &   -1.94 & 0.48   &4.18 & $4.3\times10^{13}$  & 1.184 &  $5.5\times10^{12}$ &  276 $\pm$           13 &  187 $\pm$           17 &  $4.9\times10^{12}$ &  $1.0\times10^{13}$ &   281$\pm$   13 &  184$\pm$  17 & 115\\
1049+44  &  1 &  3.00  & 0.57   & 7.22  & $7.9\times10^{13}$  & 1.132 &  $9.0\times10^{13}$ & -339 $\pm$           40 &  539 $\pm$           66 &  $7.6\times10^{13}$ &  $1.0\times10^{15}$ &   -347$\pm$   21 &  532$\pm$  59 & -3\\
1049+44  &  2 &    1.37 & 0.57   & 7.83 & $8.0\times10^{13}$  &1.131 &  $7.6\times10^{13}$ & -227 $\pm$           43 &  475 $\pm$           36 &  $4.1\times10^{13}$ &  $9.3\times10^{14}$ &  -228$\pm$   43 &  472$\pm$  35 & 24\\
1049+44  &  3* &   -0.27 & 0.65   & 3.70 & $3.8\times10^{13}$  &1.198 &  $1.1\times10^{13}$ & -300 $\pm$           39 &  273 $\pm$           38 &  $1.1\times10^{13}$ &  $2.5\times10^{14}$ &  -297$\pm$   37 &  271$\pm$  36 & -65\\
1049+44  &  4 &   -2.35 & 0.40   & 2.80 & $2.9\times10^{13}$  &1.241 &  $1.1\times10^{13}$ & -114 $\pm$           41 &  374 $\pm$            0 &  $8.4\times10^{12}$ &  $2.6\times10^{14}$ &  -122$\pm$   41 &  237$\pm$  56 & -17\\
1056+24  &  1 &    2.98 & 0.80   & 1.97 &$2.2\times10^{13}$  & 1.310 &  $1.7\times10^{13}$ & -427 $\pm$           35 &  329 $\pm$           57 &  $2.6\times10^{13}$ &  $1.2\times10^{15}$ &  -446$\pm$    33 &  332$\pm$  56 & 52\\
1056+24  &  2* &    0.45 & 0.78   & 6.89 & $7.0\times10^{13}$ &1.142 &  $5.2\times10^{13}$ & -466 $\pm$            8 &  249 $\pm$           10 &  $1.1\times10^{14}$ &  $1.6\times10^{15}$ &  -460$\pm$   7 &   294$\pm$  9 & 57\\
1056+24  &  3 &   -1.04 & 0.86   & 4.20 & $4.3\times10^{13}$ & 1.190 &  $3.2\times10^{13}$ & -392 $\pm$          190 &  403 $\pm$          207 &  $3.4\times10^{13}$ &  $1.5\times10^{15}$ &  -394$\pm$   36 &  437$\pm$  42 & -41\\
1056+24  &  4 &   -4.17 & 0.80   & 1.78 & $1.8\times10^{13}$ &1.353 &   $8.5\times10^{9}$ & \nodata  & \nodata  &  $2.1\times10^{12}$ &  $7.7\times10^{14}$ &  -467$\pm$    274 &  487$\pm$  210 & -64\\
1056+24  &  5 &   -8.34 & 0.47   & 1.89 & $2.0\times10^{13}$ &1.337 &  $9.0\times10^{12}$ & -358 $\pm$           67 &  467 $\pm$          126 &  $9.7\times10^{12}$ &  $6.1\times10^{14}$ &  -356$\pm$   58 &   388$\pm$  170 & -70\\
1056+24&  1b &    2.98 & 0.80   & 3.75 & $3.7\times10^{13}$ &1.208 &  $2.9\times10^{13}$ & -210 $\pm$            6 &  138 $\pm$           13 &  $4.2\times10^{13}$ &  $6.5\times10^{14}$ &  -211$\pm$    5 &     145$\pm$      11 & \nodata \\
1056+24&  2b* &    0.45 & 0.78   & 2.22 & $2.3\times10^{13}$ &1.298 &  $2.8\times10^{13}$ & -186 $\pm$            7 &  208 $\pm$           12 &  $7.3\times10^{13}$ &  $7.3\times10^{14}$ &  -194$\pm$   6 &     242$\pm$     9 & \nodata \\
1056+24&  3b &   -1.04 & 0.86   & 3.39 & $3.5\times10^{13}$ &1.216 &  $3.6\times10^{13}$ & -151 $\pm$            3 &  169 $\pm$           8 &  $5.7\times10^{13}$ &  $8.2\times10^{14}$ &  -150$\pm$   2 &     191$\pm$      11 & \nodata \\
1056+24&  4b &   -4.17 & 0.80   & 2.44 & $2.5\times10^{13}$ &1.284 &  $2.3\times10^{13}$ & -181 $\pm$            5 &  125 $\pm$            9 &  $4.6\times10^{13}$ &  $6.7\times10^{14}$ &  -181$\pm$   4 &     153$\pm$      12 & \nodata \\
1056+24&  5b &   -8.34 & 0.47   & 2.91 &$3.0\times10^{13}$  &1.246 &  $1.2\times10^{13}$ & -113 $\pm$            7 &  148 $\pm$           7 &  $1.1\times10^{13}$ &  $3.0\times10^{14}$ &  -108$\pm$   4 &     163$\pm$      14 & \nodata \\
1150+13  &  1 &   5.51 & \nodata   & \nodata & \nodata   &\nodata &  \nodata & \nodata  & \nodata  &  $2.3\times10^{12}$ &  $9.4\times10^{13}$ &   66$\pm$   77 &  285$\pm$ 103 & 22\\
1150+13  &  2 &    4.05 & 0.21   & 4.22 &$4.3\times10^{13}$  &1.179 &  $1.0\times10^{13}$ &   63 $\pm$           30 &  245 $\pm$           34 &  $9.0\times10^{12}$ &  $1.9\times10^{14}$ &   56$\pm$   32 &  247$\pm$  36 & -89\\
1150+13  &  3 &    1.53 & 0.22   & 2.12 & $2.2\times10^{13}$  &1.293 &  $1.7\times10^{12}$ &  139 $\pm$           74 &  221 $\pm$            0 &  $5.4\times10^{12}$ &  $2.6\times10^{14}$ &  -43$\pm$   55 &  433$\pm$  94 & 73\\
1150+13  &  4* &   -0.15 & 0.43   & 3.10 &$3.2\times10^{13}$  &1.219 &  $2.2\times10^{13}$ &  -89 $\pm$           3.5 &  221 $\pm$            4 &  $2.7\times10^{13}$ &  $5.2\times10^{14}$ &  -90$\pm$   3 &  224$\pm$  4 & 8\\
1150+13  &  5 &  -1.84 & 0.33   & 2.72 & $2.8\times10^{13}$  &1.243 &  $1.0\times10^{13}$ & -104 $\pm$           14 &  157 $\pm$           19 &  $1.1\times10^{13}$ &  $3.0\times10^{14}$ & -103$\pm$   14 &  160$\pm$  18 & -35\\
1524+10  &  1 &    3.61 & 0.59   & 2.74 & $2.8\times10^{13}$ &1.247 &  $1.7\times10^{13}$ & -200 $\pm$            3 &  138 $\pm$            7 &  $1.4\times10^{13}$ &  $2.5\times10^{13}$ &  -201$\pm$   2 &  127$\pm$  3 & -144\\
1524+10  &  2 &    1.52 & 0.61   & 6.67 & $6.8\times10^{13}$ &1.142 &  $6.7\times10^{13}$ & -215 $\pm$            5 &  312 $\pm$           12 &  $3.2\times10^{13}$ &  $5.1\times10^{13}$ &  -211$\pm$   5 &  297$\pm$  7 & -77\\
1524+10  &  3* &  -0.72 & 0.78   & 7.11 & $7.3\times10^{13}$  &1.138 &  $9.9\times10^{13}$ & -245 $\pm$            1 &  310 $\pm$            4 &  $8.0\times10^{13}$ &  $2.6\times10^{14}$ &  -247$\pm$   3 &    307$\pm$  5 & -75\\
1524+10  &  4 &   -2.35 & 0.84   & 5.56 &$5.6\times10^{13}$  &1.157 &  $7.3\times10^{13}$ & -195 $\pm$            4 &  260 $\pm$           13 &  $6.5\times10^{13}$ &  $4.1\times10^{14}$ &  -194$\pm$    2 &   217$\pm$  5 & -42\\
1524+10  &  5 &   -4.14 & 0.75   & 3.85 & $3.9\times10^{13}$ &1.194 &  $4.0\times10^{13}$ & -186 $\pm$            7 &  218 $\pm$           16 &  $3.0\times10^{13}$ &  $7.5\times10^{14}$ &  -188$\pm$   5 &  183$\pm$  9 & -43\\
1609-01  &  1 &   3.37 & 0.58   & 2.75 & $2.8\times10^{13}$ &1.239 &  $1.3\times10^{13}$ & -314 $\pm$           25 &  132 $\pm$           32 &  $3.6\times10^{12}$ &  $6.9\times10^{12}$ &  -328$\pm$   22 &  116$\pm$  31 & -4\\
1609-01  &  2 &    1.68 & 0.43   & 2.35 &$2.4\times10^{13}$  &1.271 &  $1.8\times10^{13}$ & -207 $\pm$          112 &  409 $\pm$          121 &  $1.4\times10^{13}$ &  $4.9\times10^{14}$ &  -234$\pm$   67 &  371$\pm$   82 & -22\\
1609-01  &  3* &     0  & 0.25   & 7.82 & $8.0\times10^{13}$  &1.128 &  $1.30\times10^{13}$ & -319 $\pm$          114 &  407 $\pm$          143 &  $1.7\times10^{12}$ &  $2.1\times10^{12}$ &  -308$\pm$   37 &   338$\pm$  58 & 6\\
1609-01  &  4 &   -1.68 &  \nodata  & \nodata & \nodata & \nodata &        \nodata & \nodata  & \nodata  &  $1.4\times10^{12}$ &  $1.7\times10^{12}$ &  -230$\pm$   88 &   409$\pm$  319 & 87\\
1609-01  &  5 &   -4.13 & \nodata   & \nodata & \nodata &\nodata &        \nodata & \nodata  & \nodata &  $1.7\times10^{12}$ &  $9.1\times10^{13}$ &  (-386)   &  210$\pm$  94 & 10\\
1720-00  &  1 &    1.63 & 0.58   & 5.59 & $5.7\times10^{13}$  & 1.157 &  $4.1\times10^{13}$ & -451 $\pm$            5 &  213 $\pm$            8 &  $1.7\times10^{14}$ &  $6.7\times10^{14}$ &  -450$\pm$   4 &  212$\pm$   7 & -121\\
1720-00  &  2* &    0   & 0.70   & 7.78 & $7.9\times10^{13}$  &1.135 &  $8.0\times10^{13}$ & -397 $\pm$            6 &  352 $\pm$           12 &  $4.4\times10^{14}$ &  $1.0\times10^{15}$ &  -402$\pm$   5 &  342$\pm$ 10 & -31\\
1720-00  &  3 &   -2.39 & 0.55   & 2.90 & $3.0\times10^{13}$  &1.244 &  $1.8\times10^{13}$ & -281 $\pm$            4 &  217 $\pm$            7 &  $1.1\times10^{14}$ &  $5.1\times10^{14}$ &  -281$\pm$    4 &  223$\pm$  7 & 110\\
1720-00  &  4 &   -4.03 & 0.61   &  10.4 & $1.1\times10^{14}$ &1.120 &  $1.2\times10^{14}$ & -280 $\pm$           60 &  533 $\pm$           73 &  $6.1\times10^{13}$ &  $1.3\times10^{15}$ &  -290$\pm$   43 &  565$\pm$  58 & 79\\
1720-00  &  5 &   -5.52 & 0.77   & 9.00 & $8.9\times10^{13}$ &1.128 &  $1.2\times10^{14}$ & -188 $\pm$           12 &  407 $\pm$           24 &  $1.9\times10^{14}$ &  $1.4\times10^{15}$ &  -196$\pm$   11 &  412$\pm$  24 & 164\\
1836+35  &  1 &    5.66 & \nodata   & \nodata & \nodata & \nodata &        \nodata & \nodata  & \nodata  &  $2.4\times10^{11}$ &  $1.6\times10^{13}$ &  -259$\pm$   299 &  664$\pm$  1516 & -91\\
1836+35  &  2 &    1.68 & 0.55   & 6.01 & $5.9\times10^{13}$  &1.151 &  $6.0\times10^{13}$ & -400 $\pm$           11 &  384 $\pm$           25 &  $1.5\times10^{13}$ &  $1.0\times10^{15}$ &  -420$\pm$   37 &  375$\pm$  40 & 63\\
1836+35  &  3* &    0   &  0.62   & 3.71  & $3.8\times10^{13}$  &1.196 &  $2.2\times10^{13}$ & -348 $\pm$           56 &  407 $\pm$           29 &  $1.2\times10^{13}$ &  $1.0\times10^{15}$ &  -347$\pm$   158 &  395$\pm$  175 & -200\\
1836+35  &  4 &   -1.68 & 0.46   &  2.18 &$2.8\times10^{13}$  &1.240 &  $7.4\times10^{12}$ & -280 $\pm$           51 &  202 $\pm$           46 &  $6.9\times10^{12}$ &  $7.0\times10^{14}$ &  -259$\pm$   70 &  244$\pm$  43 & -183\\
1836+35  &  5 &   -4.59 & \nodata   & \nodata & \nodata &\nodata &        \nodata & \nodata  & \nodata  &  $1.8\times10^{12}$ &  $1.4\times10^{14}$ &  -342$\pm$   294 &  347$\pm$  282 & -36\\
1836+35&  1b &    5.66 &  \nodata  & \nodata &  \nodata & \nodata &        \nodata & \nodata  & \nodata  &     $1.6\times10^{12}$ &  $8.4\times10^{13}$  &   34$\pm$   15 &     (89) & \nodata\\
1836+35&  2b &    1.68 &  0.55   & 2.16 & $2.4\times10^{13}$ &1.272 &  $2.3\times10^{12}$ & -135 $\pm$           33 &   (89) &    $5.8\times10^{12}$ & $3.1\times10^{14}$  &  -315$\pm$   65 &     345$\pm$61 & \nodata\\
1836+35&  3b* &    0   &   0.62  & 2.915 & $3.3\times10^{13}$ &1.226 &  $2.3\times10^{13}$ & -120 $\pm$            7 &  226 $\pm$           15 &    $3.5\times10^{13}$ & $5.6\times10^{14}$   &  -121$\pm$   24 &     222$\pm$6 & \nodata\\
1836+35&  4b &   -1.68 &  0.46   & 4.60 &$4.1\times10^{13}$  &1.185 &  $2.4\times10^{13}$ &  -82 $\pm$           21 &  242 $\pm$           11 &    $1.2\times10^{13}$ & $4.0\times10^{14}$  &  -71$\pm$   23 &     227$\pm$18 & \nodata\\
1836+35&  5b &   -4.59 & \nodata   & \nodata & \nodata & \nodata &        \nodata & \nodata  & \nodata  &      $2.2\times10^{12}$ &      $1.2\times10^{14}$ & -10$\pm$   23 &     181$\pm$ 43 & \nodata\\
1929-04  &  1 &    2.0 & 0.33   & 2.59 & $2.6\times10^{13}$ &1.259 &  $3.6\times10^{12}$ & -389 $\pm$           24 &  191 $\pm$           30 &  $3.3\times10^{12}$ &  $1.8\times10^{14}$ &  -391$\pm$   25 &  192$\pm$  31 & -179\\
1929-04  &  2 &    1.0 & 0.38   & 9.19 & $9.4\times10^{13}$ &1.123 &  $3.1\times10^{13}$ & -411 $\pm$           38 &  441 $\pm$           55 &  $1.8\times10^{13}$ &  $6.7\times10^{14}$ &  -405$\pm$   42 &  456$\pm$  65 & -126\\
1929-04  &  3* &  -0.5 & 0.25   & 4.39 & $4.5\times10^{13}$  &1.177 &  $1.8\times10^{13}$ & -403 $\pm$           28 &  321 $\pm$           50 &  $8.5\times10^{12}$ &  $3.1\times10^{14}$ &   -399$\pm$   31 &  352$\pm$  68 & -57\\
2008-03  &  1 &   1.87 & \nodata   & 2.06 &$2.1\times10^{13}$  &1.314 &  $1.0\times10^{13}$ & -360 $\pm$           12 &  193 $\pm$           14 &  $3.2\times10^{13}$ &  $8.5\times10^{14}$ &  -225$\pm$   68 &  575$\pm$  79 & 50\\
2008-03  &  2* &   0.18 & 0.52   & 9.95 & $1.0\times10^{14}$ & 1.120 &  $1.2\times10^{14}$ & -419 $\pm$            0 &  651 $\pm$           21 &  $2.7\times10^{13}$ &  $1.1\times10^{15}$ &  -398$\pm$   44 &  680$\pm$  60 & -31\\
2008-03  &  3 &  -1.49 & 0.53   & 9.50 & $9.7\times10^{13}$ &1.121 &  $9.8\times10^{13}$ & -391 $\pm$           15 &  651 $\pm$           39 &  $1.1\times10^{14}$ &  $1.0\times10^{15}$ &  -396$\pm$   14 &  649$\pm$  39 & -126\\
2008-03  &  4 &  -3.18 & 0.42   & 7.42 & $7.6\times10^{13}$  &1.135 &  $4.8\times10^{13}$ & -404 $\pm$           29 &  526 $\pm$           75 &  $3.5\times10^{13}$ &  $6.1\times10^{14}$ &  -411$\pm$   29 &   525$\pm$  74 & -82\\
2336+36  &  1 &    3.56 & 0.10   & 2.45 & $2.5\times10^{13}$ &1.277 &        \nodata & \nodata  & \nodata   &  $1.6\times10^{12}$ &  $8.6\times10^{13}$ &  (-396)     &  (328) & -41\\
2336+36  &  2 &    1.92 & 0.22   & 3.56 & $3.6\times10^{13}$ & 1.210 &  $9.9\times10^{12}$ & -478 $\pm$           63 &  (328) &  $2.3\times10^{12}$ &  $3.1\times10^{12}$ &  (-396) &  284$\pm$  61 & 23\\
2336+36  &  3* &    0.28 &  0.28  & 7.44 & $7.6\times10^{13}$ &1.136 &  $2.6\times10^{13}$ & -417 $\pm$           23 &  295 $\pm$           28 &  $2.0\times10^{13}$ &  $3.2\times10^{14}$ &  -413$\pm$   26 &  299$\pm$  30 & -19\\
2336+36  &  4 &   -1.36 & 0.43   & 3.04 & $3.1\times10^{13}$ &1.234 &  $1.2\times10^{13}$ & -300 $\pm$          158 &  460 $\pm$          269 &  $9.6\times10^{12}$ &  $2.6\times10^{14}$ &  -381$\pm$   56 &  303$\pm$  66 & -115\\
2336+36  &  5 &   -2.99 & 0.22   & 2.15 & $2.2\times10^{13}$ &1.304 &  $6.2\times10^{12}$ & -370 $\pm$          191 &  458 $\pm$          344 &  $3.6\times10^{12}$ &  $1.8\times10^{14}$ &  -345$\pm$   239 &   376$\pm$  438 & -19\\
2336+36  &  6 &  -6.57 & \nodata   & \nodata & \nodata & \nodata &        \nodata & \nodata  & \nodata  &  $2.5\times10^{12}$ &  $1.3\times10^{14}$ &  -747$\pm$   1081 &  399$\pm$   3408 & \nodata \\
\enddata
\\
\tablecomments{
(1) Galaxy;
(2) Aperture number from Figure~1, where asterisk denotes aperture nearest the R-band nucleus;
(3) Offset of aperture from nucleus;
(4) Estimated covering factor;
(5) Fitted optical depth of the dynamic component;
(6) Column density of the dynamic component from optical depth fit in col (5);
(7) Doublet ratio chosen for intensity fit based on value in column (6);
(8) Column density of the dynamic component from intensity fit with doublet ratio $D_R$ from column (7);
(9) Doppler shift of the dynamic component (relative to CO velocity) from the intensity fit with $D_R$ from column (7);
(10)FWHM of the dynamic component from the intensity fit with $D_R$ from column (7);
(11) Maximum column density from the intensity fit with variable $D_R$, where 
the maximum of columns (11), (6), and (8) should be taken as the best value;
(12) Minimum column density from the intensity fit with variable $D_R$,
the minimum of columns (12),  (6),  and (8) should be takend as the best value;
(13) V from the intensity fit with variable $D_R$;
(14) FWHM from the intensity fit with variable $D_R$.
Throughout this table, parameters listed in parentheses were held constant to get an adequate fit. 
}
\label{tab:data}  \end{deluxetable}


Complete text and figures available in a single document
at:~~http://www.physics.ucsb.edu/\~cmartin/publications.html.
Figures 1a-o and Figure 11 available as GIF files on astro-ph.




  \begin{figure}
         \caption{	   ({\it Top Left}) Position
	   of the apertures on an R band image from Murphy \et (1996).
	   North is up, and east is to the left.
	   The slit  is 20\asec long.
	   ({\it Top Right}) The \Ha plus [NII] $\lambda \lambda 6548, 6584$
	   emission lines along the slit; wavelength increases to the right. Spectral apertures
	   and slit orientation are indicated. 
	   ({\it Bottom Right })  Extracted \Ha spectra with apertures and slit orientation marked.
	   ({\it Bottom Left}) Extracted spectra at \nad $\lambda \lambda  5890,96$.
	   The nuclear spectrum (red), aperture number, and slit orientation are marked.
	   The blue line indicates the systemic velocity of the $\lambda 5896$~line.  
	   Two \nad\  doublets are plotted ; the velocity of one is fixed at the 
	   systemic velocity (dotted), and the velocity of the dynamic component is fitted
	   (solid).
	   The line shapes of two galaxies required a minimum of two dynamic components, and
	   the second dynamic component is shown by a dashed doublet. 
	   The emission line is \ion{He}{1} $\lambda 5876$. 
	   (a) \i10565+2448 -   Blueshifted, interstellar absorption is detected  everywhere 
	   continuum is detected.  In contrast, the \Ha emission is redshifted (blueshifted)
	   west (east) of the nucleus consistent with galactic rotation. The companion to 
	   the southeast shows no \Ha emission (i.e. little star formation) but is
	   apparently covered by the outflow.  The \nad\ linewidth and covering factor
	   decrease away from the nucleus, but the average outflow velocity remains near
	   300\kms.
	   (b) \i17208-0014 -  The \nad\ absorption is highly blueshifted
	    across both of the merging galaxies; the outflow speed is
	    400\kms at the nucleus.  The \Ha emission reveals rotation across the
	  main galaxy and the prograde nature of the encounter.  In contrast to
	  the \i10565+2448 spectrum, the \nad\ absorption shows the same velocity 
	  gradient as the \Ha emission here.  Additional galaxies are shown in 
	  the electronic edition as follows:
	  (c) \i00153+5454,  (d) \i00188-0856, (e) \i00262+4251, (f) \i03158+4227,
	  (g) \i08030+5243, (h) \i10494+4424, (i) \i11506+1331, (j) \i15245+1019,
	  (k) \i16090-0139, (l) \i16487+5447, (m) \i18368+3549, (n) \i19297-0406,
	  (o) \i20087-0308, (p) \i23365+3604.
}
\label{fig:nad_spec} 
\end{figure}

\clearpage
  \begin{figure}[h]

\vspace{10cm}

FIGURE 2a-o available as f2a-o.gif 

      \caption{Gas velocities measured across the ULIGs in \Ha (circles) and 
	\nad.	(squares). Apertures are marked in the top left panel of Figure~1.
	The \Ha emission traces the projected rotation of the merging galaxy 
	pair along the slit.  The \nad\ absorption traces the speed of
	the cool outflow. Aperture numbers, slit orientations, and the
	location of double galaxies are marked for comparison to Figure 1. 
	The galaxies are: (a) \i10565+2448, (b) \i17208-0014, (c) \i00153+5454,
	(d) \i00188-0856, and (e) \i00262+4251, (f) \i03158+4227,
	(g) \i08030+5243, (h) \i10494+4424, (i) \i11506+1331, (j) \i15245+1019,
	(k) \i16090-0139, (l) \i18368+3549, (m) \i19297-0406, (n) \i20087-0308,
	(o) \i23365+3604.
	Note that two {\it dynamic} absorption components were fitted in 
	\i10565+2448  and \i18368+3549 as discussed in Martin (2005), so
	there are two \nad\ measurements at each position.  Uncertainties
	in the \Ha points are described in \S 3.2.1.
	}
        \label{fig:pv} \end{figure}


\clearpage
  \begin{figure}[h]
\vspace{10cm}

AVAILABLE AS f3.gif

          \caption{Comparison of \nad\ and \Ha\ velocity gradients along the slit.
	    The total \nad\ velocity gradient is compared to the \Ha\ velocity
	    gradient: (Left) across the entire system; (Center)  across 
	    one galaxy -- i.e. galactic rotation; (Right) between the nuclei, i.e.
	    orbital motion.  The small solid squares denote  single nuclei systems.
	    The velocity shear across the outflowing gas is comparable to
	    that of the bulk interstellar medium, which reflects galactic and
	    orbital rotation.
	  }
           \label{fig:velgrad} \end{figure}

\clearpage
  \begin{figure}[h]
\vspace{10cm}

AVAILABLE AS f4.gif

          \caption{Position-velocity diagrams for \nad\ absorption grouped
	    by velocity gradient.
	    (a) The absorption velocity is nearly constant along
	    the slit. If the outflow diverges from the central starburst
	    region, these gradients are consistent with projection along 
	    the major axis of the gaseous disk.  
	    (b) The absorption velocity jumps near the starburst nucleus.
	    The conical wind model shows this type of shear when projected
	    along the minor axis of the gaseous disk.
	    (c) The velocity of the absorbing gas presents rotation.  
	    Rotation is not consistent with a flow diverging from
	    a nuclear starburst.  A larger launch region and/or additional 
	    torques are indicated.
	  }
          \label{fig:vgrad_na} \end{figure}

%
%

\clearpage
  \begin{figure}[h]

\vspace{10cm}

FIGURE 5 available as f5.gif 

          \caption{Masses of the cool outflows estimated by
	    integrating the \nad\ column density over each galaxy.
	    The non-linear scaling of the \nad column density with 
	    the line equivalent width creates a large uncertainty 
	    in any one mass estimate. The columns were estimated from 
	    the apparent optical depth and could be much higher if
	    strongly saturated components are present.  
	    The conversion from N(\ion{Na}{1}) to total gas column 
	    is assumed to be a factor of $ \approx 10$ (see text for details). 
	    The results suggest  $ \sgreat\ 10^{8}$\msun of cool gas is carried
	    by ULIG winds. The small  range of far-infrared 
	    luminosity indicates the baseline in star formation rates among 
	    the systems shown is small. The symbol denotes
	    the dynamical age of the merger:  double nuclei 
	    (solid), single nuclei with prominent tidal tails (open), and
	    single nuclei (crosses). 
                }
         \label{fig:m}  \end{figure}

\clearpage
  \begin{figure}[h]

\vspace{10cm}

FIGURE 6 available as f6.gif 

         \caption{Mass of the cool outflow vs starburst dynamical
	   mass. Outflow mass is from Table~\ref{tab:nad}. Dynamical
	   mass estimated from CO linewidths by Solomon \et (1997).
	   Triangles denote wind masses from
	   Schwartz \& Martin (2004) and Schwartz (2005), molecular
	   gas masses  for M~82 (Weiss \et 2001) and \n1614 
	   (Scoville \et 1991),  atomic gas mass from 
	   Kobulnicky \et (1999) for He2-10, and the atomic gas
	   masses tabulated in Schwartz \& Martin for the remaining
	   systems.  Taking the dynamical mass in the starburst region as 
	   a measure of the total gas mass, the diagonal lines
	   show outflow gas fractions of 100\% (dotted), 10\% (dashed),
	   and 1\% (solid).
	 }
	 \label{fig:mdyn} \end{figure}

\clearpage
  \begin{figure}[h]

\vspace{10cm}

FIGURE 7 available as f7.gif 

          \caption{Color excess versus \ion{Na}{1} equivalent width.
	    Uncertainties are dominated by corrections for stellar
	    absorption; and the full range allowed is shown by
	    the error bars.  ULIGs with poorly constrained reddening
	    are shown by open squares. 
	    Filled circles are LIGs from Heckman \et (2000).  The
	    systems with stronger interstellar \nad\ tend to be
	    more reddened.
	    }
         \label{fig:bv}  \end{figure}

\clearpage
  \begin{figure}[h]

\vspace{10cm}

FIGURE 8 available as f8.gif 

         \caption{Covering  fraction of absorbers vs. logarithmic 
	   extinction at \Hb. Data are nuclear measurements for the ULIGs 
	   (squares) and the LIRGs from Heckman \et (2000) (circles). Galaxies
	   with poorly constrained reddening are shown by open 
	   symbols.  Systems with nearly black absorption lines are 
	   highly reddened.
	   }
	 \label{fig:bvcf} \end{figure}


\clearpage
  \begin{figure}[h]
\vspace{10cm}

AVAILABLE AS f9.gif

         \caption{Estimated kinetic energy of the cool outflow vs
	   feedback energy supplied by 10~Myr starburst.
	   The lines denote  100\% (dotted), 10\% (dashed),
	   and 1\% (solid) of the energy supplied by a 10~Myr old
	   starburst.  Triangles denote dwarf starbursts from
	   Schwartz \& Martin (2004), who used a larger corretion
	   for sodium ionization and depletion than assumed here
	   for the ULIGs (see text for details).
	 }
	 \label{fig:ke} \end{figure}

  \begin{figure}[h]

\vspace{10cm}

FIGURE 10 available as f10.gif 

          \caption{Mass loss efficiency in cool phase vs terminal outflow speed. 
	    Estimates of the  mass flux are normalized by 
	    the star formation rate (of ~1 to 100\msun\ stars), and the
	    velocities of the most blue-shifted absorption are estimated from
	    Paper I for ULIGs (square points). The supershell mass from
	    Schwartz \& Martin (2004) was divided by a dynamical age of 10~Myr
	    to estimate mass flux (triangles).
	    The dotted lines illustrate models for energy-driven winds in
	    which  100, 10, or 1\% of the mechanical power is thermalized.  
	    Dashed lines describe the shallower relations for momentum-driven
	    winds in which 100, 10, or 1\% of the mechanical momentum goes
	    into the wind. The available mechanical power and momentum (per unit SFR)
	    is taken from a Starburst~99 
	    population synthesis model with continuous star formation for 10~Myr, 
	    solar metallicity, a Salpeter initial mass function, and lower (upper) 
	    mass  limits of 1\msun\ (100\msun).  The ULIG data points (squares)
	    are consistent with outflows driven by a few percent of the 
	    mechanical energy or $\sim 10$\% of the momentum.  The variation
	    in mass-loss  efficiency with terminal velocity distinguish these models;
	    but this interpretation is subject to how the dwarf starburst points
	    are estimated.  
                }
          \label{fig:mm} \end{figure}

\clearpage
  \begin{figure}[h]
\vspace{10cm}

\vspace{10cm}

AVAILABLE AS f11.gif

          \caption{Geometrical toy model of a conical wind leaving a nuclear
	    starburst region of radius $R_0$.  The rotational velocity of 
	    the galactic disk, $V_c$, and the radial outflow velocity, $V_w$,
	    are projected onto the observer's detector at upper right.
	    The observer is positioned at infinity in the yz-plane. The
	    disk is inclined by an angle $\iota$ with respect to the observer.
	    The red sightline intersects the major axis of the disk, which
	    is aligned with the x-axis in the diagram.
	    }
         \label{fig:toymodel}  \end{figure}

\clearpage

\vspace{10cm}

FIGURE 12a,b available as f12a,b.gif 

  \begin{figure}[h]
          \caption{Projected (rotation plus outflow) velocity of toy
	    model for constant-$L$, wind (red) and constant-$\Omega$, 
	    wind (blue).
	    The radial wind speed is taken as twice the rotation speed
	    of the disk; both are constants. The half-opening angle of the
	    wind cone is 60\deg. The thick-solid, solid, and dashed lines
	    denote viewing angles of 55\deg, 35\deg, and 15\deg, respectively.
	    There is no wind inside the launch radius, $R_{0}$.
	    (a) {\it Galaxy major axis:} 
	    The step function shows the rotation curve of a disk 
	    rotating at $V_c$. Points mark the starburst ring at
	    $R_{0}$, where the wind is launched (green). 
	    Compared to the disk, the wind presents little 
	    velocity gradient regardless of viewing angle.
	    (b) {\it Galaxy minor axis:}
	    No rotation is observed along the minor axis. The wind cone
	    produces a velocity jump because sightlines on the near-side
	    of the disk are more aligned with the outflow than those
	    on the far side, which are nearly perpendicular to it (red lines).
	    }
         \label{fig:vx3}  \end{figure}

\end{document}